\documentclass[journal,onecolumn,11pt]{IEEEtran}
\pdfoutput=1
\usepackage{graphicx} 
\graphicspath{{.}}
\usepackage[caption=false,font=footnotesize]{subfig}

\usepackage{xspace}
\usepackage{amsmath,amssymb,amsfonts}
\usepackage{xcolor} 
\usepackage{multirow} 
\usepackage{booktabs} 
\usepackage[normalem]{ulem} 
\usepackage{url} 
\usepackage[colorlinks=true]{hyperref} 
\usepackage{balance}
\usepackage{wrapfig}

\newcommand{\reals}{\ensuremath{\mathbb{R}}}

\newcommand{\mat}[1]{\ensuremath{\mathbf{#1}}}
\newcommand{\st}{\ensuremath{\,\mathrm{s.t.}\,}}
\newcommand{\norm}[1]{\ensuremath{\left\|#1\right\|}}
\newcommand{\cost}[1]{\ensuremath{\ell_{#1}}}
\newcommand{\abs}[1]{\ensuremath{\left|#1\right|}}
\newcommand{\setdef}[1]{\ensuremath{\left\{#1\right\}}}

\newcommand{\opt}[1]{{#1}^*} \newcommand{\refeq}[1]{(\ref{#1})}

\newcommand{\col}[1]{_{#1}}
\newcommand{\row}[1]{^{T}_{#1}}
\newcommand{\iter}[1]{^{(#1)}}

\newcommand{\ndims}{M}
\newcommand{\natoms}{K}
\newcommand{\nsamples}{N}
\def\actset{\mathcal{A}}

\def\modelclass{\ensuremath{\mathcal{M}}}
\def\complexity{\ensuremath{\mathcal{C}}}

\def\datam{\mat{X}\xspace} 
\def\dictm{\mat{D}\xspace}
\def\coefm{\mat{A}\xspace}

\def\datav{\mat{x}\xspace} 
\def\dictv{\mat{d}\xspace}
\def\coefv{\mat{a}\xspace}

\def\coef{a}

\def\reg{\psi}
\def\tmin{\theta_{1}}
\def\tmax{\theta_{2}}

\newcommand{\psnr}{\textsc{psnr}\xspace}
\newcommand{\moe}{\textsc{moe}\xspace}
\newcommand{\cmoe}{\textsc{cmoe}\xspace}
\newcommand{\joe}{\textsc{joe}\xspace}

\newcommand{\omp}{\textsc{omp}\xspace}

\newcommand{\mle}{\textsc{mle}\xspace}

\definecolor{NewTextBG}{rgb}{0.8,0.8,1.00}

\def\eliminar#1{{\color{red}}}

\newcommand{\best}[1]{{\bf\color{blue}#1}\xspace}

\title{%
Universal Regularizers For Robust Sparse Coding and Modeling%
}

 \author{%
 \IEEEauthorblockN{Ignacio Ram\'{i}rez and Guillermo Sapiro}\\
 \IEEEauthorblockA{%
Department of Electrical and Computer Engineering\\%
University of Minnesota\\%
\{ramir048,guille\}@umn.edu}
 }

\begin{document}
\maketitle

\begin{abstract}
  Sparse data models, where data is assumed to be well represented as a
  linear combination of a few elements from a dictionary, have gained
  considerable attention in recent years, and their use has led to
  state-of-the-art results in many signal and image processing tasks. It is
  now well understood that the choice of the sparsity regularization term is
  critical in the success of such models. 
  Based on a codelength minimization interpretation of sparse
    coding, and using tools from universal coding theory, we propose %
  a framework for designing sparsity regularization terms which have 
  theoretical and practical advantages when compared to the more standard
  $\cost{0}$ or $\cost{1}$ ones.  The presentation of the framework and
  theoretical foundations is complemented with examples that show its
practical advantages  in image denoising,
  zooming and classification.
\end{abstract}

\section{Introduction}
\label{sec:intro}
\emph{Sparse modeling} calls for constructing a succinct representation of
some data as a combination of a few typical patterns (\emph{atoms}) learned from
the data itself. Significant contributions to the theory and practice of
learning such collections of atoms (usually called \emph{dictionaries} or
\emph{codebooks}), e.g., \cite{aharon06,engan00,olshausen97}, and of representing
the actual data in terms of them, e.g., \cite{chen98,daubechies04,efron04},
have been developed in recent years, leading to state-of-the-art results in
many signal and image processing tasks
\cite{krishnapuram04,mairal08e,mairal08b,raina07}. We refer the
reader for example to \cite{bruckstein09} for a recent review on the
subject.

A critical component of sparse modeling is the actual sparsity of the
representation, which is controlled by a regularization term
(\emph{regularizer} for short) and its associated parameters.  The choice of
the functional form of the regularizer and its parameters is a challenging
task. Several solutions to this problem have been proposed in the
literature, ranging from the automatic tuning of the
parameters~\cite{giryes08} to Bayesian models, where these parameters are
themselves considered as random variables
\cite{figueiredo01,giryes08,zou06}. In this work we adopt the interpretation
of sparse coding as a codelength minimization problem. This is a natural and
objective method for assessing the quality of a statistical model for
describing given data, and which is based on the Minimum Description Length
(MDL) principle \cite{rissanen84}. In this framework, the regularization
term in the sparse coding formulation is interpreted as the cost in bits of
describing the sparse linear coefficients used to reconstruct the data.
Several works on image coding using this approach were developed in the
1990's under the name of ``complexity-based'' or ``compression-based''
coding, following the popularization of MDL as a powerful statistical
modeling tool \cite{coifman92,moulin99,saito94}. The focus on these early
works was in denoising using wavelet basis, using either generic asymptotic
results from MDL or fixed probability models, in order to compute the
description length of the coefficients.  A later, major breakthrough in MDL
theory was the adoption of \emph{universal coding} tools to compute optimal
codelengths. In this work, we improve and extend on previous results in this
line of work by designing regularization terms based on such universal codes
for image coefficients, meaning that the codelengths obtained when encoding
the coefficients of any (natural) image with such codes will be close to the
shortest codelengths that can be obtained with any model fitted specifically
for that particular instance of coefficients. The resulting framework not
only formalizes sparse coding from the MDL and universal coding perspectives
but also leads to a family of \emph{universal regularizers} which we show to
consistently improve results in image processing tasks such as denoising and
classification. These models also enjoy several desirable theoretical and
practical properties such as statistical consistency (in certain cases),
improved robustness to outliers in the data, and improved sparse signal
recovery (e.g., decoding of sparse signals from a compressive sensing point
of view \cite{candes06}) when compared with the traditional \cost{0} and
\cost{1}-based techniques in practice. %
These models also yield to the use of a simple and efficient optimization
technique for solving the corresponding sparse coding problems as a series
of weighted \cost{1} subproblems, which in turn, can be solved with off-the-shelf
algorithms such as LARS \cite{efron04} or IST \cite{daubechies04}. Details
are given in the sequel.

Finally, we apply our universal regularizers not only for coding using fixed
dictionaries, but also for learning the dictionaries themselves, leading to
further improvements in all the aforementioned tasks.

The remainder of this paper is organized as follows: in
Section~\ref{sec:sparse-modeling} we introduce the standard framework of
sparse modeling. Section~\ref{sec:universal-sparse-coding} is dedicated to
the derivation of our proposed universal sparse  modeling framework, while
Section~\ref{sec:implementation} deals with its
implementation. Section~\ref{sec:results} presents experimental results
showing the practical benefits of the proposed framework in image denoising,
zooming and classification tasks. Concluding remarks are given in
Section~\ref{sec:conclusions}.

%
\section{Sparse modeling and the need for better models}
%
\label{sec:sparse-modeling}

Let $\datam \in \reals^{\ndims{\times}\nsamples}$ be a set of $\nsamples$
column data samples $\datav\col{j} \in \reals^\ndims$, $\dictm \in
\reals^{\ndims{\times}\natoms}$ a dictionary of $\natoms$ column atoms
$\dictv\col{k} \in \reals^\ndims$, and $\coefm \in
\reals^{\natoms{\times}\nsamples}, \coefv\col{j} \in \reals^\natoms$, a set
of reconstruction coefficients such that $\datam=\dictm\,\coefm$. We use
$\coefv\row{k}$ to denote the $k$-th row of $\coefm$, 
the coefficients associated to the $k$-th atom in $\dictm$.  For each
$j=1,\ldots,\nsamples$ we define the \emph{active set} of $\coefv\col{j}$ as
$\actset_{j}=\setdef{k: \coef_{kj} \neq 0, 1 \leq k \leq \natoms}$, and
$\norm{\coefv\col{j}}_0=|\actset_{j}|$ as its cardinality.  The goal of
sparse modeling is to design a dictionary $\dictm$ such that for all or most
data samples $\datav\col{j}$, there exists a coefficients vector
$\coefv\col{j}$ such that $\datav\col{j}\approx\dictm\,\coefv\col{j}$ and
$\norm{\coefv\col{j}}_0$ is small (usually below some threshold $L \ll
\natoms$).  Formally, we would like to solve the following problem
%
\begin{align}
\min_{\dictm,\coefm} &\; \sum_{j=1}^{\nsamples}  \reg(\coefv\col{j}) 
\quad\st\;\norm{\datav\col{j} - \dictm\,\coefv\col{j}}_2^2 \leq \epsilon,\quad j=1,\ldots,\nsamples,
\label{eq:sparse-modeling}
\end{align}
where $\reg(\cdot)$ is a regularization term which induces sparsity in the
columns of the solution $\coefm$. Usually the constraint
$\norm{\dictv\col{k}}_2 \leq 1,\;k=1,\ldots,\natoms$, is added, since
otherwise we can always decrease the cost function arbitrarily by
multiplying $\dictm$ by a large constant and dividing $\coefm$ by the same
constant.  When $\dictm$ is fixed, the problem of finding a sparse
$\coefv\col{j}$ for each sample $\datav\col{j}$ is called sparse coding,
\begin{align}
\coefv\col{j} = \arg\min_{\coefv}&\;\reg(\coefv\col{j})\quad
\st\quad \norm{\datav\col{j} - \dictm\,\coefv\col{j}}_2^2 \leq \epsilon.
\label{eq:sparse-coding}
\end{align}

Among possible choices of $\reg(\cdot)$ are the \cost{0} pseudo-norm,
$\reg(\cdot)=\norm{\cdot}_0$, and the \cost{1} norm. The former tries to
solve directly for the sparsest $\coefv\col{j}$, but since it is non-convex,
it is commonly replaced by the \cost{1} norm, which is its closest convex
approximation. Furthermore, under certain conditions on (fixed) \dictm and
the sparsity of $\coefv\col{j}$, the solutions to the \cost{0} and
\cost{1}-based sparse coding problems coincide (see for example
\cite{candes06}). The problem \refeq{eq:sparse-modeling} is also usually
formulated in Lagrangian form,
\begin{equation}
\min_{\dictm,\coefm}\;\sum_{j=1}^{\nsamples}{\norm{
\datav\col{j} - \dictm\,\coefv\col{j}}_2^2 + \lambda\reg(\coefv\col{j})
},
\label{eq:sparse-modeling-lagrangian}
\end{equation}
\noindent along with its respective sparse coding problem when $\dictm$ is
fixed,
\begin{equation}
  \coefv\col{j} = \arg\min_{\coefv}\; \norm{\datav\col{j} - 
  \dictm\,\coefv}_2^2 + \lambda\reg(\coefv).
  \label{eq:sparse-coding-lagrangian}
\end{equation}
Even when the regularizer $\reg(\cdot)$ is convex, the sparse modeling
problem, in any of its forms, is jointly non-convex in
$(\dictm,\coefm)$. Therefore, the standard approach to find an approximate
solution is to use alternate minimization: starting with an initial
dictionary $\dictm\iter{0}$, we minimize
\refeq{eq:sparse-modeling-lagrangian} alternatively in $\coefm$ via
\refeq{eq:sparse-coding} or \refeq{eq:sparse-coding-lagrangian} (sparse
coding step), and $\dictm$ (dictionary update step). The sparse coding step
can be solved efficiently when $\reg(\cdot)=\norm{\cdot}_1$ using for
example \textsc{ist} \cite{daubechies04} or \textsc{lars}~\cite{efron04}, or
with \textsc{omp}~\cite{mallat93} when $\reg(\cdot)=\norm{\cdot}_0$. The
dictionary update step can be done using for example
\textsc{mod}~\cite{engan00} or \textsc{k-svd}~\cite{aharon06}.

\subsection{Interpretations of the sparse coding problem}
\label{sec:sparse-coding:interpretation}

We now turn our attention to the sparse coding problem: given a fixed
dictionary $\dictm$, for each sample vector $\datav\col{j}$, compute the
sparsest vector of coefficients $\coefv\col{j}$ that yields a good
approximation of $\datav\col{j}$. The sparse coding problem admits several
interpretations. What follows is a summary of these interpretations and the
insights that they provide into the properties of the sparse models that are
relevant to our derivation.
\subsubsection{Model selection in statistics} 
\label{sec:sparse-coding:interpretation:statistics}
Using the \cost{0} norm as
$\reg(\cdot)$ in \refeq{eq:sparse-coding-lagrangian} is known in the
statistics community as the Akaike's Information Criterion (\textsc{aic})
when $\lambda=1$, or the Bayes Information Criterion (\textsc{bic}) when
$\lambda=\frac{1}{2}\log \ndims$, two popular forms of model selection (see
\cite[Chapter 7]{hastie08}). In this context, the \cost{1} regularizer was
introduced in \cite{tibshirani96}, again as a convex approximation of the
above model selection methods, and is commonly known (either in its
constrained or Lagrangian forms) as the \emph{Lasso}.  Note however that, in the
regression interpretation of \refeq{eq:sparse-coding-lagrangian}, the
role of $\dictm$ and $\datam$ is very different.
%

\subsubsection{Maximum a posteriori} 

Another interpretation of
\refeq{eq:sparse-coding-lagrangian} is that of a maximum a posteriori
(\textsc{map}) estimation of $\coefv\col{j}$ in the logarithmic scale, that
is
\begin{eqnarray}
\coefv\col{j}  &=& \arg\max_{\coefv}\;\{\log P(\coefv|\datav\col{j})\} 
= \arg\max_{\coefv}\;\{\log P(\datav\col{j}|\coefv)+\log P(\coefv)\} \nonumber\\
              &=& \arg\min_{\coefv}\;\{-\log P(\datav\col{j}|\coefv)-\log P(\coefv)\},
\label{eq:map-gen}
\end{eqnarray}
where the observed samples $\datav\col{j}$ are assumed to be contaminated
with additive, zero mean, \textsc{iid} Gaussian noise with variance
$\sigma^2$,
%
$P(\datav\col{j}|\coefv) \propto e^{-\frac{1}{2\sigma^2}\norm{\datav\col{j}-\dictm\,\coefv}_2^2},$
%
\noindent and a \emph{prior probability model} on $\coefv$ with the form
%
$P(\coefv) \propto e^{-\theta\reg(\coefv)}$
%
\noindent is considered. The energy term in Equation
\refeq{eq:sparse-coding-lagrangian} follows by plugging the previous two
probability models into \refeq{eq:map-gen} and factorizing $2\sigma^2$ into
$\lambda=2 \sigma^2 \theta$.  According to \refeq{eq:map-gen}, the \cost{1}
regularizer corresponds to an \textsc{iid} Laplacian prior with mean $0$ and
inverse-scale parameter $\theta$, $P(\coefv) =
\prod_{k=1}^{\natoms}{\theta}e^{-\theta|\coef_k|} = \theta^\natoms
e^{-\theta\norm{\coefv}_1}$, which has a special meaning in signal
processing tasks such as image or audio compression. This is due to the
widely accepted fact that representation coefficients derived from
predictive coding of continuous-valued signals, and, more generally,
responses from zero-mean filters, are well modeled using Laplacian
distributions. For example, for the special case of \textsc{dct}
coefficients of image patches, an analytical study of this phenomenon is
provided in \cite{lam00}, along with further references on the subject.%

\subsubsection{Codelength minimization}
\label{sec:sparse-coding:interpretation:codelength} Sparse coding, in all
its forms, has yet another important interpretation. Suppose that we have a
fixed dictionary $\dictm$ and that we want to use it to compress an image,
either losslessly by encoding the reconstruction coefficients $\coefm$ and
the residual $\datam-\dictm\coefm$, or in a lossy manner, by obtaining a
good approximation $\datam \approx \dictm\coefm$ and encoding only
$\coefm$. Consider for example the latter case. Most modern compression
schemes consist of two parts: a probability assignment stage, where the
data, in this case $\coefm$, is assigned a probability $P(\coefm)$, and an
encoding stage, where a code $C(\coefm)$ of length $L(\coefm)$ bits is
assigned to the data given its probability, so that $L(\coefm)$ is as short
as possible.  The techniques known as Arithmetic and Huffman coding provide
the best possible solution for the encoding step, which is to approximate
the Shannon ideal codelength $L(\coefm)=-\log P(\coefm)$
\cite[Chapter~5]{cover06}. Therefore, modern compression theory deals with
finding the coefficients $\coefm$ that maximize $P(\coefm)$, or, equivalently,
that minimize $-\log P(\coefm)$. Now, to encode $\datam$ lossily, we obtain
coefficients $\coefm$ such that each data sample $\datav\col{j}$ is
approximated up to a certain \cost{2} distortion $\epsilon$,
$\norm{\datav\col{j}-\dictm\coefv\col{j}}_2^2 \leq \epsilon$. Therefore, given
a model $P(\coefv)$ for a vector of reconstruction coefficients, and
assuming that we encode each sample independently, the optimum vector of
coefficients $\coefv\col{j}$ for each sample $\datav\col{j}$ will be the
solution to the optimization problem
\begin{equation}
\coefv\col{j} = \arg\min_{\coefv}\;-\log P(\coefv)\quad
\st\quad       \norm{\datav\col{j} - \dictm\,\coefv\col{j}}_2^2 \leq \epsilon,
\label{eq:lossy-encoding}
\end{equation}
\noindent which, for the choice $P(\coefv) \propto e^{-\reg(\coefv)}$
coincides with the error constrained sparse coding problem
\refeq{eq:sparse-coding}.  Suppose now that we want lossless compression. In
this case we also need to encode the reconstruction residual
$\datav\col{j}-\dictm\coefv\col{j}$. Since $P(\datav,\coefv) = P(\datav|\coefv)P(\coefv)$,
the combined codelength will be
\begin{equation}
  L(\datav\col{j},\coefv\col{j}) 
  = -\log P(\datav\col{j},\coefv\col{j}) 
  = -\log P(\datav\col{j}|\coefv\col{j}) - \log P(\coefv\col{j}).
\label{eq:codelength}
\end{equation}
Therefore, obtaining the best coefficients $\coefv\col{j}$ amounts to solving
$\min_{\coefv}L(\datav\col{j},\coefv\col{j})$, which is precisely the
\textsc{map} formulation of \refeq{eq:map-gen}, which in turn, for proper
choices of $P(\datav|\coefv)$ and $P(\coefv)$, leads to the Lagrangian form
of sparse coding \refeq{eq:sparse-coding-lagrangian}.\footnote{%
  Laplacian models, as well as Gaussian models, are probability
  distributions over $\reals$, characterized by continuous probability
  density functions, $f(\coef)=F'(\coef)$, $F(\coef)=P(x \leq \coef)$. If
  the reconstruction coefficients are considered real numbers, under any of
  these distributions, any instance of $\coefm \in
  \reals^{\natoms\times\nsamples}$ will have measure $0$, that is,
  $P(\coefm)=0$. In order to use such distributions as our models for the
  data, we assume that the coefficients in $\coefm$ are quantized to a
  precision $\Delta$, small enough for the density function $f(\coef)$ to be
  approximately constant in any interval
  $[\coef-\Delta/2,\coef+\Delta/2],\,\coef \in \reals$, so that we can
  approximate $P(\coef) \approx \Delta f(\coef),\; \coef \in \reals$. Under
  these assumptions, $-\log P(\coef) \approx -\log f(\coef) -\log \Delta$,
  and the effect of $\Delta$ on the codelength produced by any model is the
  same.  Therefore, we will omit $\Delta$ in the sequel, and treat density
  functions and probability distributions interchangeably as $P(\cdot)$. Of
  course, in real compression applications, $\Delta$ needs to be tuned. }

As one can see, the codelength interpretation of sparse coding is able to
unify and interpret both the constrained and unconstrained formulations into
one consistent framework. Furthermore, this framework offers a natural and
objective measure for comparing the quality of different models
$P(\datav|\coefv)$ and $P(\coefv)$ in terms of the codelengths obtained.

\subsubsection{Remarks on related work}
As mentioned in the introduction, the codelength interpretation of signal
coding was already studied in the context of orthogonal wavelet-based
denoising. An early example of this line of work considers a regularization
term which uses the Shannon Entropy function $\sum p_i \log p_i$ to give a
measure of the sparsity of the solution \cite{coifman92}. However, the
Entropy function is not used as measure of the ideal codelength for
describing the coefficients, but as a measure of the sparsity (actually,
group sparsity) of the solution.  The MDL principle was applied to the
signal estimation problem in \cite{saito94}.  In this case, the codelength
term includes the description of both the location and the magnitude of the
nonzero coefficients. Although a pioneering effort, the model assumed in
\cite{saito94} for the coefficient magnitude is a uniform distribution on
$[0,1]$, which does not exploit a priori knowledge of image coefficient
statistics, and the description of the support is slightly
wasteful. Furthermore, the codelength expression used is an asymptotic
result, actually equivalent to \textsc{bic} (see
Section~\ref{sec:sparse-coding:interpretation:statistics}) which can be
misleading when working with small sample sizes, such as when encoding small
image patches, as in current state of the art image processing applications.
The uniform distribution was later replaced by the \emph{universal code for
  integers} \cite{rissanen92} in \cite{moulin99}. However, as in
\cite{saito94}, the model is so general that it does not perform well for
the specific case of coefficients arising from image decompositions, leading
to poor results. In contrast, our models are derived following a careful
analysis of image coefficient statistics.  Finally, probability models
suitable to image coefficient statistics of the form $P(\coef) \propto
e^{-|\coef|^{\beta}}$ (known as generalized Gaussians) were applied to the
MDL-based signal coding and estimation framework in \cite{moulin99}. The
justification for such models is based on the empirical observation that
sparse coefficients statistics exhibit ``heavy tails'' (see next
section). However, the choice is ad hoc and no optimality criterion is
available to compare it with other possibilities. Moreover, there is no closed
form solution for performing parameter estimation on such family of models,
requiring numerical optimization techniques.  In
Section~\ref{sec:universal-sparse-coding}, we derive a number of probability
models for which parameter estimation can be computed efficiently in closed
form, and which are guaranteed to optimally describe image coefficients.

\subsection{The need for a better model}
\label{sec:sparse-coding:improvements}

\begin{figure*}
\begin{center}
\includegraphics[width=6.5in]{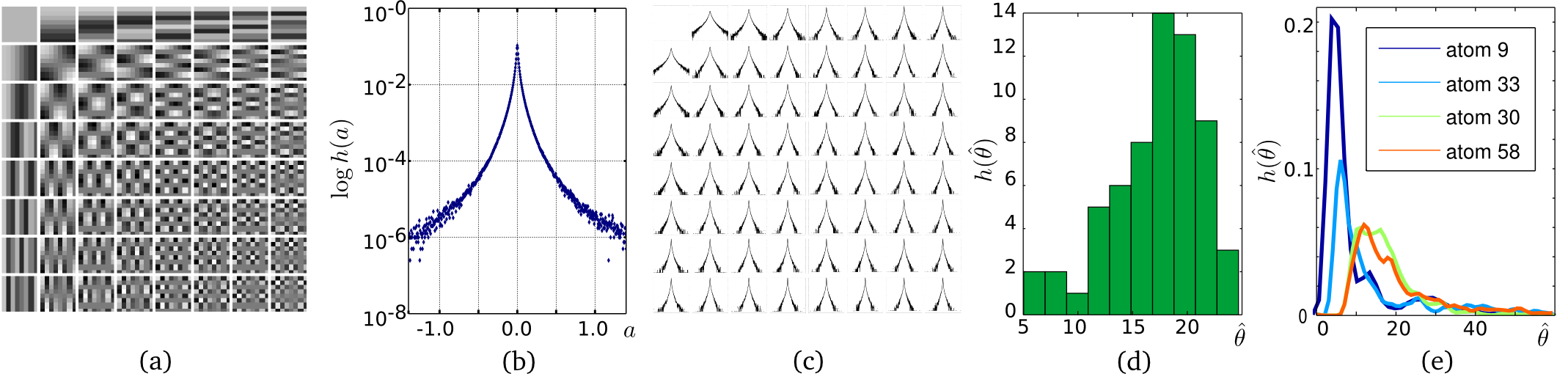}%
\caption{\label{fig:many-laplacians}\footnotesize%
  Standard $8{\times}8$ \textsc{dct} dictionary (a), global empirical
  distribution of the coefficients in $\coefm$ (b, log scale),
  empirical distributions of the coefficients associated to each of the
  $\natoms=64$ \textsc{dct} atoms (c, log scale). The distributions
    in (c) have a similar heavy tailed shape (heavier than Laplacian), but
    the variance in each case can be significantly different.
  (d) Histogram of the $\natoms=64$ different %
  $\hat\theta_k$ values obtained by fitting a Laplacian distribution to each
  row $\coefv\row{k}$ of $\coefm$. Note that there are significant
  occurrences between $\hat\theta=5$ to $\hat\theta=25$.  %
  The coefficients $\coefm$ used in (b-d) were obtained from encoding $10^6$
  $8{\times}8$ patches (after removing their DC component) randomly sampled
  from the Pascal 2006 dataset of natural images \cite{pascal06}.  (e)
  Histograms showing the spatial variability of the best local estimations
  of $\hat\theta_k$ for a few rows of $\coefm$ across different regions of
  an image. In this case, the coefficients $\coefm$ correspond to the sparse
  encoding of all $8{\times}8$ patches from a single image, in scan-line
  order. For each $k$, each value of $\hat\theta_k$ was computed from a
  random contiguous block of $250$ samples from $\coefv\row{k}$. The
  procedure was repeated $4000$ times to obtain an empirical
  distribution. The wide supports of the empirical distributions indicate
  that the estimated $\hat\theta$ can have very different values, even for
  the same atom, depending on the region of the data from where the
  coefficients are taken.%
}
  \end{center}
\end{figure*}

As explained in the previous subsection, the use of the \cost{1} regularizer
implies that all the coefficients in $\coefm$ share the same Laplacian
parameter $\theta$. However, as noted in \cite{lam00} and references
therein, the empirical variance of coefficients associated to different
atoms, that is, of the different rows $\coefv\row{k}$ of $\coefm$, varies
greatly with $k=1\,\ldots,\natoms$.  This is clearly seen in
Figures~\ref{fig:many-laplacians}(a-c), which show the empirical
distribution of \textsc{dct} coefficients of $8{\times}8$ patches. As the
variance of a Laplacian is $2/\theta^2$, different variances indicate
different underlying $\theta$.  The histogram of the set
$\setdef{\hat\theta_k,k=1,\ldots,\natoms}$ of estimated Laplacian parameters
for each row $k$, Figure~\ref{fig:many-laplacians}(d), shows that this is
indeed the case, with significant occurrences of values of $\hat\theta$ in a
range of $5$ to $25$.
The straightforward modification suggested by this phenomenon is to use a
model where each row of $\coefm$ has its own weight associated to it,
leading to a weighted \cost{1} regularizer. However, from a modeling
perspective, this results in $\natoms$ parameters to be adjusted instead of
just one, which often results in poor generalization properties. For
example, in the cases studied in Section~\ref{sec:results}, even with
thousands of images for learning these parameters, the results of applying
the learned model to new images were always significantly worse (over 1dB in
estimation problems) when compared to those obtained using simpler models
such as an unweighted \cost{1}. \footnote{Note that this is the case when
  the weights are found by maximum likelihood. Other applications of
  weighted \cost{1} regularizers, using other types of weighting strategies,
  are known to improve over \cost{1}-based ones for certain applications
  (see e.g. \cite{zou06}).}  One reason for this failure may be that real
images, as well as other types of signals such as audio samples, are far
from stationary. In this case, even if each atom $k$ is associated to its
own $\theta_k$ ($\lambda_k$), the optimal value of $\theta_k$ can have
significant local variations at different positions or times. This effect is
shown in Figure~\ref{fig:many-laplacians}(e), where, for each $k$, each
$\theta_k$ was re-estimated several times using samples from different
regions of an image, and the histogram of the different estimated values of
$\hat\theta_k$ was computed. Here again we used the \textsc{dct} basis as
the dictionary $\dictm$.

The need for a flexible model which at the same time has a small number of
parameters leads naturally to Bayesian formulations where the different
possible $\lambda_k$ are ``marginalized out'' by imposing an hyper-prior
distribution on $\lambda$, sampling $\lambda$ using its posterior
distribution, and then averaging the estimates obtained with the sampled
sparse-coding problems. Examples of this recent line of work, and the
closely related Bayesian Compressive Sensing, are developed for example in
\cite{ji09,tipping01,wipf07,wipf03}. Despite of its promising results, the
Bayesian approach is often criticized due to the potentially expensive
sampling process (something which can be reduced  for certain
choices of the priors involved \cite{ji09}), arbitrariness in the choice of the priors,
and lack of proper theoretical justification for the proposed models \cite{wipf03}.

In this work we pursue the same goal of deriving a more flexible and
accurate sparse model than the traditional ones, while avoiding an increase
in the number of parameters and the burden of possibly solving several
sampled instances of the sparse coding problem. For this, we deploy tools
from the very successful information-theoretic field of universal
  coding, which is an extension of the compression scenario summarized
above in Section~\ref{sec:sparse-coding:interpretation}, when the probability
model for the data to be described is itself unknown and has to be described
as well.


%
\section{Universal models for sparse coding}
%
\label{sec:universal-sparse-coding}

Following the discussion in the preceding section, we now have several
possible scenarios to deal with. First, we may still want to consider a
single value of $\theta$ to work well for all the coefficients in $\coefm$,
and try to design a sparse coding scheme that does not depend on prior
knowledge on the value of $\theta$. Secondly, we can consider an independent
(but not identically distributed) Laplacian model where the underlying
parameter $\theta$ can be different for each atom $\dictv\col{k}$,
$k=1,\ldots,\natoms$. In the most extreme
scenario, we can consider each single coefficient $\coef_{kj}$ in $\coefm$
to have its own unknown underlying $\theta_{kj}$ and yet, we would like to
encode each of these coefficients (almost) as if we knew its hidden
parameter.

The first two scenarios are the ones which fit the original purpose of
universal coding theory \cite{merhav98}, which is the design of optimal codes
for data whose probability models are unknown, and the models themselves are
to be encoded as well in the compressed representation. 

Now we develop the basic ideas and techniques of
universal coding applied to the first scenario, where the problem is to
describe $\coefm$ as an \textsc{iid} Laplacian with unknown parameter
$\theta$.  Assuming a known parametric form for the prior, with unknown
parameter $\theta$, leads to the concept of a \emph{model class}. In our
case, we consider the class $\mathcal{M}=\setdef{P(\coefm|\theta):\theta \in
  \Theta}$ of all \textsc{iid} Laplacian models over $\coefm \in
\reals^{\natoms{\times}\nsamples}$, where
\[
P(\coefm|\theta)=\prod_{j=1}^{\nsamples}\prod_{k=1}^{\natoms}P(\coef_{kj}|\theta),\quad
P(\coef_{kj}|\theta) = \theta e^{-\theta|\coef_{kj}|}
\]
and $\Theta \subseteq \reals^{+}$. The goal of universal coding is to find a
probability model $Q(\coefm)$ which can fit $\coefm$ as well as the model in
$\mathcal{M}$ that best fits $\coefm$ after having observed it. A model
$Q(\coefm)$ with this property is called \emph{universal} (with respect to
the model $\mathcal{M}$).

\def\ncoeffs{n}
For simplicity, in the following discussion we consider
the coefficient matrix $\coefm$ to be arranged as a single long column
vector of length $\ncoeffs=\natoms{\times}\nsamples$,
$\coefv=(\coef_1,\ldots,\coef_\ncoeffs)$. We also use the letter $a$ without
sub-index to denote the value of a random variable representing coefficient
values.

First we need to define a criterion for comparing the fitting quality of
different models. In universal coding theory this is done in terms of
the codelengths $L(\coefv)$ required by each model to describe $\coefv$. 

If the model consists of a single probability distribution $P(\cdot)$,
we know from Section~\ref{sec:sparse-coding:interpretation:codelength} that
the optimum codelength corresponds to $L_P(\coefv)=-\log
P(\coefv)$. Moreover, this relationship defines a one-to-one correspondence
between distributions and codelengths, so that for any coding scheme
$L_Q(\coefv)$, $Q(\coefv)=2^{-L_Q(\coefv)}$. %
Now suppose that we are restricted to a class of models $\modelclass$, and
that we need choose the model $\hat{P} \in\modelclass$ that assigns the
shortest codelength to a particular instance of $\coefv$. We then have that
$\hat{P}$ is the model in $\modelclass$ that assigns the maximum probability to
$\coefv$. For a class $\modelclass$ parametrized by $\theta$, this corresponds to
$\hat{P}=P(\coefv|\hat\theta(\coefv))$, %
where $\hat\theta(\coefv)$
is the maximum likelihood estimator (\textsc{mle}) of the model class
parameter $\theta$ given $\coefv$ (we will usually omit the argument and
just write $\hat\theta$).  Unfortunately, we also need to include the value of $\hat\theta$
in the description of $\coefv$ for the decoder to be able to reconstruct it
from the code $C(\coefv)$. Thus, we have that any model $Q(\coefv)$ inducing
valid codelengths $L_Q(\coefv)$ will have $L_Q(\coefv) > -\log
P(\coefv|\hat\theta)$. The overhead of $L_Q(\coefv)$ with respect to
$-\log P(\coefv|\hat\theta)$ is known as the \emph{codelength regret},%
\[
\mathcal{R}(\coefv,Q) := L_Q(\coefv) - (- \log P(\coefv|\hat\theta(\coefv))) =
-\log Q(\coefv) + \log P(\coefv|\hat\theta(\coefv))).
\]%
A model $Q(\coefv)$ (or, more precisely, a sequence of models, one for each
data length $\ncoeffs$) is called \emph{universal} if
$\mathcal{R}(\coefv,Q)$ grows sublinearly in $\ncoeffs$ for all possible
realizations of $\coefv$, that is
$
\frac{1}{\ncoeffs}\mathcal{R}(\coefv,Q) \rightarrow 0\,,\;\forall\,\coefv \in \reals^\ncoeffs,
$
so that the codelength regret with respect to the \textsc{mle} becomes
asymptotically negligible.  

There are a number of ways to construct
universal probability models. The simplest one is the so called
\emph{two-part code}, where the data is described in two parts. The first
part describes the optimal parameter $\hat\theta(\coefv)$ and the second
part describes the data according to the model with the value of the
estimated parameter $\hat\theta$, $P(\coefv|\hat\theta(\coefv))$. For
uncountable parameter spaces $\Theta$, such as a compact subset of $\reals$,
the value of $\hat\theta$ has to be quantized in order to be described with
a finite number of bits $d$. We call the quantized parameter
$\hat\theta_d$. The regret for this model is thus
\[
\mathcal{R}(\coefv,Q) = L(\hat\theta_d) + L(\coefv|\hat\theta_d) - L(\coefv|\hat\theta)
= L(\hat\theta_d) -\log P(\coefv|\hat\theta_d) - (-\log P(\coefv|\hat\theta)).
\]
The key for this model to be universal is in the choice of the quantization
step for the parameter $\hat\theta$, so that both its description
$L(\hat\theta_d)$, and the difference $-\log P(\coefv|\hat\theta_d) - (-\log
P(\coefv|\hat\theta))$, grow sublinearly. This can be achieved by letting
the quantization step shrink as $O(1/\sqrt{n})$ \cite{rissanen84}, thus
requiring $d=O(0.5\log n)$ bits to describe each dimension of
$\hat\theta_d$. This gives us a total regret for two-part codes which grows
as $\frac{\mathrm{dim}(\Theta)}{2}\log n$, where $\mathrm{dim}(\Theta)$ is
the dimension of the parameter space $\Theta$.

Another important universal code is the so called \emph{Normalized Maximum
Likelihood} (\textsc{nml}) \cite{shtarkov87}. In this case the universal
model $\opt{Q}(\coefv)$ corresponds to the model that minimizes the worst
case regret,
\[
\opt{Q}(\coefv) = \min_{Q} \max_{\coefv} \{-\log Q(\coefv) + \log P(\coefv|\hat\theta(\coefv))\},
\]

\noindent which can be written in closed form as $\opt{Q}(\coefv) =
\frac{P(\coefv|\hat\theta(\coefv))}{\complexity(\modelclass,n)}$, where the
normalization constant $$\complexity(\modelclass,n) := \sum_{\coefv \in
  \reals^n}{P(\coefv|\hat\theta(\coefv))d\coefv}$$ is the value of the
minimax regret and depends only on $\modelclass$ and the length of the data
$\ncoeffs$.\footnote{The minimax optimality of  $\opt{Q}(\coefv)$
derives from the fact that it defines a complete uniquely decodable code for
all data $\coefv$ of length $\ncoeffs$, that is, it satisfies the Kraft
inequality with equality.%
$\sum_{\coefv \in \reals^\ncoeffs} 2^{-L_{\opt{Q}}(\coefv)} = 1.$ Since
every uniquely decodable code with lengths $\setdef{L_Q(\coefv):\coefv \in
  \reals^n}$ must satisfy the Kraft inequality (see
\cite[Chapter~5]{cover06}), if there exists a value of $\coefv$ such that
$L_Q(\coefv) < L_{\opt{Q}}(\coefv)$ (that is $2^{-L_Q(\coefv)} >
2^{-L_{\opt{Q}}(\coefv)}$), then there exists a vector $\coefv'$ for
which $L_Q(\coefv') > L_{\opt{Q}}(\coefv')$ for the Kraft inequality to
hold.  Therefore the regret of $Q$ for $\coefv'$ is necessarily greater than
$\complexity(\modelclass,n)$, which shows that $\opt{Q}$ is minimax
optimal.} Note that the \textsc{nml} model requires
$\complexity(\modelclass,n)$ to be finite, something which is often not the
case.

The two previous examples are good for assigning a probability to
coefficients $\coefv$ that have already been computed, but they cannot be
used as a model for computing the coefficients themselves since they depend
on having observed them in the first place. For this and other reasons that
will become clearer later, we concentrate our work on a third important
family of universal codes derived from the so called \emph{mixture models}
(also called \emph{Bayesian mixtures}).  In a mixture model, $Q(\coefv)$ is
a convex mixture of all the models $P(\coefv|\theta)$ in \modelclass,
indexed by the model parameter $\theta$,
$Q(\coefv)=\int_{\Theta}{P(\coefv|\theta)w(\theta)d\theta}$, where
$w(\theta)$ specifies the weight of each model. Being a convex mixture
implies that $w(\theta)\geq0$ and $\int_\Theta w(\theta)d\theta = 1$, thus
$w(\theta)$ is itself a probability measure over $\Theta$. We will restrict
ourselves to the particular case when $\coefv$ is considered a sequence of
independent random variables,\footnote{More sophisticated models which
  include dependencies between the elements of $\coefv$ are out of the scope
  of this work.}
\begin{equation}
Q(\coefv)=\prod_{j=1}^{\ncoeffs}Q_j(\coef_j),
\quad Q_j(\coef_j)=\int_{\Theta}{P(\coef_j|\theta)w_j(\theta)d\theta},
\label{eq:model-mixture}
\end{equation}

where the mixing function $w_j(\theta)$ can be different for each sample $j$. An
important particular case of this scheme is the so called \emph{Sequential
  Bayes} code, in which $w_j(\theta)$ is computed sequentially as a
posterior distribution based on previously observed samples, that is
$w_j(\theta) = P(\theta|\coef_1,\coef_2,\ldots,\coef_{n-1})$ \cite[Chapter~6]{grunwald07}. 
In this work, for simplicity, we restrict ourselves to the case where
$w_j(\theta)=w(\theta)$ is the same for all $j$. The result is an
\textsc{iid} model where the probability of each sample $\coef_j$ is a mixture
of some probability measure over $\reals$,
\begin{equation}
Q_j(a_j) = Q(\coef_j) = \int_{\Theta}{P(\coef_j|\theta)w(\theta)d\theta},\;\forall\,j=1,\ldots,\nsamples.
\label{eq:scalar-mixture}
\end{equation}

A well known result for \textsc{iid} mixture (Bayesian) codes states that
their asymptotic regret is $O(\frac{\mathrm{dim}(\Theta)}{2}\log n)$, thus
stating their universality, as long as the weighting function $w(\theta)$ is
positive, continuous and unimodal over $\Theta$ (see for example
\cite[Theorem~8.1]{grunwald07},\cite{schwartz78}). This gives us great
flexibility on the choice of a weighting function $w(\theta)$ that
guarantees universality. Of course, the results are asymptotic and the
$o(\log n)$ terms can be large, so that the choice of $w(\theta)$ can have
practical impact for small sample sizes.

In the following discussion we derive several \textsc{iid} mixture models
for the Laplacian model class \modelclass. For this purpose, it will be
convenient to consider the corresponding one-sided counterpart of the
Laplacian, which is the exponential distribution over the absolute value of
the coefficients, $|\coef|$, and then symmetrize back to obtain the final
distribution over the signed coefficients $\coef$.

\subsection{The conjugate prior}
\label{sec:universal:conjugate}

In general, \refeq{eq:scalar-mixture} can be computed in closed form if
$w(\theta)$ is the conjugate prior of $P(\coef|\theta)$. When
$P(\coef|\theta)$ is an exponential (one-sided Laplacian), the conjugate
prior is the Gamma distribution,
\[
w(\theta|\kappa,\beta) =
  {\Gamma(\kappa)}^{-1}\theta^{\kappa-1}\beta^{\kappa}e^{-\beta\theta},\; \theta \in \reals^{+},
\]
where $\kappa$ and $\beta$ are its \textit{shape} and \textit{scale}
parameters respectively. Plugging this in \refeq{eq:scalar-mixture} we
obtain the \textit{Mixture of exponentials} model (\moe), which
has the following form (see Appendix \ref{sec:moe-details} for the full derivation),
\begin{equation}
  Q_{\moe}(\coef|\beta,\kappa) = \kappa\beta^\kappa(\coef+\beta)^{-(\kappa+1)},\;\coef \in \reals^{+}.
  \label{eq:moe-prior-one-sided}
\end{equation}
With some abuse of notation, we will also denote the symmetric distribution
on $\coef$ as \moe,
\begin{equation}
  Q_{\moe}(\coef|\beta,\kappa) = \frac{1}{2}\kappa\beta^\kappa(|\coef|+\beta)^{-(\kappa+1)},\;\coef \in \reals.
  \label{eq:moe-prior}
\end{equation}

Although the resulting prior has two parameters to deal with instead of one,
we know from universal coding theory that, in principle, any choice of
$\kappa$ and $\beta$ will give us a model whose codelength regret is
asymptotically small. 

Furthermore, being \textsc{iid} models, each coefficient of $\coefv$ itself
is modeled as a mixture of exponentials, which makes the resulting model
over $\coefv$ very well suited to the most flexible scenario where the
``underlying'' $\theta$ can be different for each $\coef_j$. In
Section~\ref{sec:results:fitting} we will show that a single \moe
distribution can fit each of the $\natoms$ rows of $\coefm$ better than
$\natoms$ separate Laplacian distributions fine-tuned to these rows, with a
total of $\natoms$ parameters to be estimated. Thus, not only  we can deal
with one single unknown $\theta$, but we can actually achieve maximum flexibility
with only two parameters ($\kappa$ and $\beta$). This property is particular
of the mixture models, and does not apply to the other universal models
presented.

Finally, if desired, both $\kappa$ and $\beta$ can be easily estimated using
the method of moments (see Appendix~\ref{sec:moe-details}). Given sample
estimates of the first and second non-central moments,
$\hat\mu_1=\frac{1}{n}\sum_{j=1}^n{|\coef_j|}$ and
$\hat\mu_2=\frac{1}{n}\sum_{j=1}^{n}{|\coef_j|^2}$, we have that
\begin{equation}
\hat\kappa = 2(\hat\mu_2-\hat\mu_1^2)/(\hat\mu_2-2\hat\mu_1^2) \quad\mathrm{and}\quad
\hat\beta  = (\hat\kappa-1)\hat\mu_1.
\label{eq:moe-parameter-estimation}
\end{equation}
When the \moe prior is plugged into \refeq{eq:map-gen} instead of the
standard Laplacian, the following new sparse coding formulation is obtained,
\begin{equation}
  \coef\col{j} = \arg\min_{\coefv}\norm{\datav\col{j} - \dictm\coefv}_2^2 +
  \lambda_\moe\sum_{k=1}^{\natoms} \log
  \left(\abs{\coef_{k}} + \beta \right),
  \label{eq:moe-sparse-coding}
\end{equation}
where $\lambda_\moe=2\sigma^2(\kappa+1)$. An example of the \moe
regularizer, and the thresholding function it induces, is shown in
Figure~\ref{fig:regularizers} (center column) for $\kappa=2.5,\beta=0.05$.
Smooth, differentiable non-convex regularizers such as the one in in
\refeq{eq:moe-sparse-coding} have become a mainstream robust alternative to
the \cost{1} norm in statistics~\cite{fan01,zou06}. Furthermore, it has been
shown that the use of such regularizers in regression leads to consistent
estimators which are able to identify the relevant variables in a regression
model (oracle property) \cite{fan01}. This is not always the case for the
\cost{1} regularizer, as was proved in \cite{zou06}. The \moe\ regularizer
has also been recently proposed in the context of compressive sensing
\cite{candes08}, where it is conjectured to be better than the \cost{1}-term
at recovering sparse signals in compressive sensing
applications.\footnote{In~\cite{candes08}, the logarithmic regularizer
  arises from approximating the \cost{0} pseudo-norm as an
  \cost{1}-normalized element-wise sum, without the insight and theoretical
  foundation here reported.} This conjecture was partially confirmed
recently for non-convex regularizers of the form
$\reg(\coefv)=\norm{\coefv}_r$ with $0 < r < 1$ in \cite{saab08,foucart08},
and for a more general family of non-convex regularizers including the one
in \refeq{eq:moe-sparse-coding} in \cite{trzasko10}. In all cases, it was
shown that the conditions on the sensing matrix (here $\dictm$) can be
significantly relaxed to guarantee exact recovery if non-convex regularizers
are used instead of the \cost{1} norm, provided that the exact solution to
the non-convex optimization problem can be computed. In practice, this
regularizer is being used with success in a number of applications here and
in \cite{chartrand09,trzasko09}.\footnote{While these works support the use
  of such non-convex regularizers, none of them formally derives them using
  the universal coding framework as in this paper.%
} %
Our experimental results in Section~\ref{sec:results} provide further
evidence on the benefits of the use of non-convex regularizers, leading to a
much improved recovery accuracy of sparse coefficients compared to \cost{1}
and \cost{0}. We also show in Section~\ref{sec:results} that the \moe prior
is much more accurate than the standard Laplacian to model the distribution
of reconstruction coefficients drawn from a large database of image
patches. We also show in Section~\ref{sec:results} how these improvements
lead to better results in applications such as image estimation and
classification.

\subsection{The Jeffreys prior}
\label{sec:universal:jeffreys}

The Jeffreys prior for a parametric model class $\modelclass=\setdef{P(\coef|\theta),\,\theta \in
\Theta}$, is defined as
\begin{equation}
w(\theta) = \frac{\sqrt{|I(\theta)|}}{\int_{\Theta}{\sqrt{|I(\xi)|}d\xi}},\;\theta\in\Theta,
\label{eq:jeffreys-prior}
\end{equation}

\noindent where $|I(\theta)|$ is the determinant of the \emph{Fisher
  information matrix}
\begin{equation}
I(\theta)=\left.\left\{ E_{P(\coef|\tilde\theta)}
\left[-\frac{\partial^2}{\partial\tilde\theta^2}\log
  P(\coef|\tilde\theta)\right]\right\}\right|_{\tilde\theta=\theta}.
\label{eq:fisher-information}
\end{equation}

The Jeffreys prior is well known in Bayesian theory due to three important
properties: it virtually eliminates the hyper-parameters of the model, it is
invariant to the original parametrization of the distribution, and it is a
``non-informative prior,'' meaning that it represents well the lack of prior
information on the unknown parameter $\theta$ \cite{bernardo94}. It turns
out that, for quite different reasons, the Jeffreys prior is also of
paramount importance in the theory of universal coding. For instance, it has
been shown in \cite{barron98} that the worst case regret of the mixture code
obtained using the Jeffreys prior approaches that of the \textsc{nml} as the
number of samples $n$ grows. Thus, by using Jeffreys, one can attain the
minimum worst case regret asymptotically, while retaining the advantages of
a mixture (not needing hindsight of $\coefv$), which in our case means to
be able to use it as a model for computing $\coefv$ via sparse coding.

For the exponential distribution we have that
$I(\theta)=\frac{1}{\theta^2}$.  Clearly, if we let $\Theta=(0,\infty)$, the
integral in \refeq{eq:jeffreys-prior} evaluates to $\infty$. Therefore,
in order to obtain a proper integral, we need to exclude $0$ and $\infty$ from
$\Theta$ (note that this was not needed for the conjugate prior). We choose
to define $\Theta=[\tmin,\tmax]$, $0 < \tmin < \tmax < \infty$, leading to
$
w(\theta)=\frac{1}{\ln(\tmax/\tmin)}\frac{1}{\theta},\;\theta \in [\tmin,\tmax].
$

\noindent The resulting mixture, after being symmetrized around $0$, has the
following form (see Appendix \ref{sec:joe-details}):
\begin{equation}
  Q_{\joe}(\coef|\tmin,\tmax) = \frac{1}{2\ln(\tmax/\tmin)} \frac{1}{|\coef|} \left( e^{-\tmin|\coef|} -e^{-\tmax|\coef|} \right),\;\coef \in \reals^{+}.
\label{eq:joe}
\end{equation}
We refer to this prior as a \emph{Jeffreys mixture of exponentials} (\joe), and
again overload this acronym to refer to the symmetric case as well. Note
that although $Q_{\joe}$ is not defined for $\coef=0$, its limit when $\coef
\rightarrow 0$ is finite and evaluates to
$\frac{\tmax-\tmin}{2\ln(\tmax/\tmin)}$. Thus, by defining
$Q_{\joe}(0)=\frac{\tmax-\tmin}{2\ln(\tmax/\tmin)}$, we obtain a prior that
is well defined and continuous for all $\coef \in \reals$. When plugged into \refeq{eq:map-gen},
we get the \joe-based sparse coding formulation,
\begin{equation}
\min_{\coefv} \norm{\datav\col{j}-\dictm\coefv}_2^2 + 
\lambda_\joe\sum_{k=1}^{\natoms}\{\log |\coef_k| - \log (e^{-\tmin|\coef_k|} -e^{-\tmax|\coef_k|}) \},
\label{eq:sparse-coding-joe}
\end{equation}

\noindent where, according to the convention just defined for $Q_\joe(0)$,
we define $\reg_\joe(0) := \log (\tmax-\tmin)$. According
to the \textsc{map} interpretation we have that $\lambda_\joe=2\sigma^2$,
coming from the Gaussian assumption on the approximation error as explained
in Section~\ref{sec:sparse-coding:interpretation}. 

\begin{figure*}
\begin{center}
\includegraphics[width=6.5in]{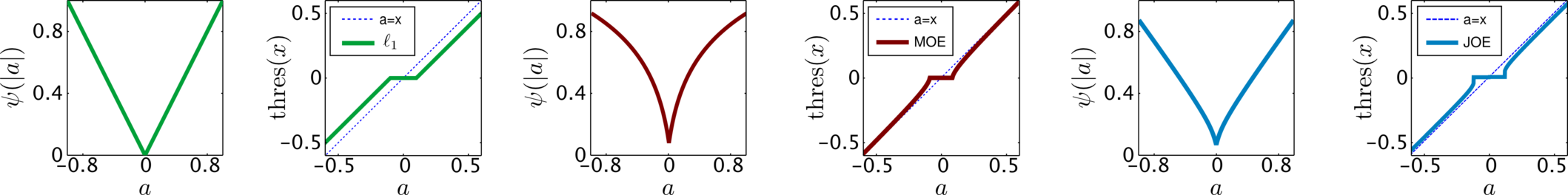} %
\caption{\label{fig:regularizers}\footnotesize%
  Left to right: \cost{1} (green), \moe (red) and \joe (blue) regularizers
  and their corresponding thresholding functions $\mathrm{thres}(x) :=
  \arg\min_\coef\{(x-\coef)^2 + \lambda\reg(|\coef|)\}$. The
  unbiasedness of \moe is due to the fact that large coefficients are not
  shrank by the thresholding function. Also, although the \joe regularizer
  is biased, the shrinkage of large coefficients can be much smaller than
  the one applied to small coefficients. }
\end{center}
\end{figure*}

As with \moe, the \joe-based regularizer, $\reg_\joe(\cdot) = -\log
Q_\joe(\cdot)$, is continuous and differentiable in $\reals^+$, and its
derivative converges to a finite value at zero, $\lim_{\coef\rightarrow
  0}\reg_\joe'(\coef) = \frac{\tmax^2-\tmin^2}{\tmax-\tmin}$. As we will see
later in Section~\ref{sec:implementation}, these properties are important to
guarantee the convergence of sparse coding algorithms using non-convex
priors. Note from \refeq{eq:sparse-coding-joe} that we can rewrite the \joe
regularizer as
\[
\reg_\joe(\coef_k) = \log |\coef_k| - 
\log e^{ -\tmin|\coef| }(1-e^{ -(\tmax-\tmin)|\coef| }) = 
\tmin |\coef_k| + \log |\coef_k| - \log ( 1 - e^{-(\tmax-\tmin)|\coef_k|} ),
\]
so that for sufficiently large $|\coef_k|$, $\log ( 1 - e^{-(\tmax-\tmin)|\coef_k|})
\approx 0$, $\tmin|\coef_k| \gg \log|\coef_k|$, and we have that
$\reg_\joe(|\coef_k|) \approx \tmin|\coef_k|$.  Thus, for large $|\coef_k|$,
the \joe regularizer behaves like \cost{1} with
$\lambda'=2\sigma^2\tmin$. In terms of the probability model, this means
that the tails of the \joe mixture behave like a Laplacian with
$\theta=\tmin$, with the region where this happens determined by the value
of $\tmax-\tmin$. %
The fact that the non-convex region of $\reg_\joe(\cdot)$ is confined to a
neighborhood around $0$ could help to avoid falling in bad local minima
during the optimization (see Section~\ref{sec:implementation} for more
details on the optimization aspects). Finally, although having Laplacian
tails means that the estimated $\coefv$ will be biased \cite{fan01}, the
sharper peak at $0$ allows us to perform a more aggressive thresholding of
small values, without excessively clipping large coefficients, which leads
to the typical over-smoothing of signals recovered using an \cost{1}
regularizer.  See Figure~\ref{fig:regularizers} (rightmost column) for an
example regularizer based on \joe with parameters $\tmin=20,\tmax=100$, and
the thresholding function it induces.

The \joe regularizer has two hyper-parameters $(\tmin,\tmax)$ which define
$\Theta$ and that, in principle, need to be tuned. One possibility is to
choose $\tmin$ and $\tmax$ based on the physical properties of the data to
be modeled, so that the possible values of $\theta$ never fall outside of
the range $[\tmin,\tmax]$. For example, in modeling patches from grayscale
images with a limited dynamic range of $[0,255]$ in a \textsc{dct} basis,
the maximum variance of the coefficients can never exceed $128^2$. The same
is true for the minimum variance, which is defined by the quantization
noise.

Having said this, in practice it is advantageous to adjust $[\tmin,\tmax]$
to the data at hand. In this case, although no closed form solutions exist
for estimating $[\tmin,\tmax]$ using \textsc{mle} or the method of
moments, standard optimization techniques can be easily applied to obtain
them. See Appendix~\ref{sec:joe-details} for details.


\subsection{The conditional Jeffreys}
\label{sec:universal:cond-jeffreys}

A recent approach to deal with the case when the integral over $\Theta$ in
the Jeffreys prior is improper, is the \emph{conditional Jeffreys}
\cite[Chapter 11]{grunwald07}. The idea is to construct a proper prior,
based on the improper Jeffreys prior and the first few $n_0$ samples of
$\coefv$, $(\coef_1,\coef_2,\ldots,\coef_{n_0})$, and then use it for the remaining
data. The key observation is that although the normalizing integral
$\int{\sqrt{I(\theta)}d\theta}$ in the Jeffreys prior is improper, the
unnormalized prior $w(\theta)=\sqrt{I(\theta)}$ can be used as a measure to
weight $P(\coef_1,\coef_2,\ldots,\coef_{n_0}|\theta)$,
\begin{equation}
  w(\theta) =
  \frac{P(\coef_1,\coef_2,\ldots,\coef_{n_0}|\theta)\sqrt{I(\theta)}}
  {\int_{\Theta}{P(\coef_1,\coef_2,\ldots,\coef_{n_0}|\xi)\sqrt{I(\xi)}d\xi}}.
\label{eq:cmoe-gen}
\end{equation}

It turns out that the integral in \refeq{eq:cmoe-gen} usually becomes proper
for small $n_0$ in the order of $\dim(\Theta)$. In our case we have that for
any $n_0 \geq 1$, the resulting prior is a
$\mathrm{Gamma}(\kappa_0,\beta_0)$ distribution with $\kappa_0 := n_0$ and
$\beta_0 := \sum_{j=1}^{n_0}{\coef_j}$ (see Appendix~\ref{sec:cmoe-details}
for details). Therefore, using the conditional Jeffreys prior in the mixture
leads to a particular instance of \textsc{moe}, which we denote by
\textsc{cmoe} (although the functional form is identical to \textsc{moe}),
where the Gamma parameters $\kappa$ and $\beta$ are automatically selected 
from the data. This may explain in part why the Gamma prior performs so well
in practice, as we will see in Section~\ref{sec:results}.

Furthermore, we observe that the value of $\beta$ obtained with this
approach ($\beta_0$) coincides with the one estimated using the method of
moments for \textsc{moe} if the $\kappa$ in \moe is fixed to
$\kappa=\kappa_0+1=n_0+1$. Indeed, if computed from $n_0$ samples, the
method of moments for \moe gives $\beta=(\kappa-1)\mu_1$, with
$\mu_1=\frac{1}{n_0}\sum{\coef_j}$, which gives us
$\beta=\frac{n_0+1-1}{n_0}\sum{\coef_j}=\beta_0$. It turns out in practice
that the value of $\kappa$ estimated using the method of moments gives a
value between $2$ and $3$ for the type of data that we deal with (see
Section~\ref{sec:results}), which is just above the minimum acceptable value
for the \cmoe prior to be defined, which is $n_0=1$. This justifies our
choice of $n_0=2$ when applying \cmoe in practice.

As $n_0$ becomes large, so does $\kappa_0=n_0$, and the Gamma prior
$w(\theta)$ obtained with this method converges to a Kronecker delta at the
mean value of the Gamma distribution,
$\delta_{\kappa_0/\beta_0}(\cdot)$. Consequently, when $w(\theta) \approx
\delta_{\kappa_0/\beta_0}(\theta)$, the mixture
$\int_\Theta{P(\coef|\theta)w(\theta)d\theta}$ will be close to
$P(\coef|\kappa_0/\beta_0)$. Moreover, from the definition of $\kappa_0$ and
$\beta_0$ we have that $\kappa_0/\beta_0$ is exactly the \textsc{mle} of
$\theta$ for the Laplacian distribution. Thus, for large $n_0$, the
conditional Jeffreys method approaches the \textsc{mle} Laplacian model.

Although from a universal coding point of view this is not a problem, for
large $n_0$ the conditional Jeffreys model will loose its flexibility to
deal with the case when different coefficients in $\coefm$ have different
underlying $\theta$. On the other hand, a small $n_0$ can lead to a prior
$w(\theta)$ that is overfitted to the local properties of the first
samples, which for non-stationary data such as image patches, can be
problematic. Ultimately, $n_0$ defines a trade-off between the degree of
flexibility and the accuracy of the resulting model.

\section{Optimization and implementation details}
\label{sec:implementation}

All of the mixture models discussed so far yield non-convex regularizers,
rendering the sparse coding problem non-convex in $\coefv$.  It turns out
however that these regularizers satisfy certain conditions which make the
resulting sparse coding optimization well suited to be approximated using a
sequence of successive convex sparse coding problems, a technique known as
\emph{Local Linear Approximation} (\textsc{lla}) \cite{zou08} (see also
\cite{trzasko09,gasso09} for alternative optimization techniques for such
non-convex sparse coding problems). In a nutshell, suppose we need to obtain
an approximate solution to
\begin{equation}
  \coefv\col{j} = \arg\min_{\coefv} \norm{\datav\col{j} - 
  \dictm\,\coefv}_2^2 + \lambda\sum_{k=1}^{\natoms}\reg(|\coef_k|),
  \label{eq:sparse-coding-lagrangian-bis}
\end{equation}
where $\reg(\cdot)$ is a  non-convex function over $\reals^{+}$. At each
\textsc{lla} iteration, we compute $\coefv\col{j}\iter{t+1}$ by doing a first
order expansion of $\reg(\cdot)$ around the $\natoms$ elements of the current estimate
$\coef_{kj}\iter{t}$,
\[
\tilde\reg_k\iter{t}(|\coef|) = \reg(|\coef_{kj}\iter{t}|) + 
\reg'(|\coef_{kj}\iter{t}|)\left( |\coef| - |\coef_{kj}\iter{t}| \right)
= \reg'(|\coef_{kj}\iter{t}|)|\coef| + c_k,
\]

\noindent and solving the convex weighted \cost{1} problem that results
after discarding the constant terms $c_k$,
\begin{eqnarray}
  \coefv\col{j}\iter{t+1} 
  &=& \arg\min_{\coefv} \norm{ \datav\col{j} - \dictm\,\coefv}_2^2 + 
    \lambda\sum_{k=1}^{\natoms}{\tilde\reg_k\iter{t}(|\coef_k|)}\nonumber\\
  &=& \arg\min_{\coefv} \norm{ \datav\col{j} - \dictm\,\coefv}_2^2 + 
\lambda\sum_{k=1}^{\natoms}{\reg'(|\coef_{kj}\iter{t}|)|\coef_k|}
  = \arg\min_{\coefv} \norm{ \datav\col{j} - \dictm\,\coefv}_2^2 + 
  \sum_{k=1}^{\natoms}{\lambda_k\iter{t}|\coef_k|}.
  \label{eq:lla-sparse-coding}
\end{eqnarray}
\noindent where we have defined $\lambda_k\iter{t}
:=\lambda\reg'(|\coef_{kj}\iter{t}|)$. %
If $\reg'(\cdot)$ is continuous in $(0,+\infty)$,
and right-continuous and finite at $0$, then the \textsc{lla} algorithm
converges to a stationary point of \refeq{eq:sparse-coding-lagrangian-bis}
\cite{zou06}. These conditions are met for both the \moe and \joe
regularizers. Although, for the \joe\ prior, the derivative $\reg'(\cdot)$
is not defined at $0$, it converges to the limit
$\frac{\tmax^2-\tmin^2}{2(\tmax-\tmin)}$ when $|\coef| \rightarrow 0$, which
is well defined for $\tmax \neq \tmin$. If $\tmax=\tmin$, the \joe\ mixing
function is a Kronecker delta and the prior becomes a Laplacian with
parameter $\theta=\tmin=\tmax$. Therefore we have that for all of the
mixture models studied, the \textsc{lla} method converges to a stationary
point. In practice, we have observed that $5$ iterations are enough to
converge. Thus, the cost of sparse coding, with the proposed non-convex
regularizers, is at most $5$ times that of a single \cost{1} sparse coding,
and could be less in practice if warm restarts are used to begin each
iteration.

\noindent Of course we need a starting point $\coefv\col{j}\iter{0}$, and,
being a non-convex problem, this choice will influence the approximation
that we obtain. One reasonable choice, used in this work, is to define
$\coef_{kj}\iter{0}=\coef_0,\,k=1,\ldots,\natoms,j=1,\ldots,\nsamples$,
where $\coef_0$ is a scalar so that $\reg'(\coef_0)=E_{w}[\theta]$, that is,
so that the first sparse coding corresponds to a Laplacian regularizer whose
parameter is the average value of $\theta$ as given by the mixing prior
$w(\theta)$.

Finally, note that although the discussion here has revolved around the
Lagrangian formulation to sparse coding of~(\ref{eq:sparse-coding-lagrangian}), this technique is
also applicable to the constrained formulation of sparse-coding given by
Equation \refeq{eq:sparse-modeling} for a fixed dictionary $\dictm$.

{\bf Expected approximation error}: Since we are solving a convex
approximation to the actual target optimization problem, it is of interest
to know how good this approximation is in terms of the original cost
function. To give an idea of this, after an approximate solution $\coefv$ is
obtained, we compute the expected value of the difference between the true
and approximate regularization term values. The expectation is taken,
naturally, in terms of the assumed distribution of the coefficients in
$\coefv$. Since the regularizers are separable, we can compute the error in
a separable way as an expectation over each $k$-th coefficient,
$\zeta_q(\coef_k)=E_{\nu \sim q}\left[ \tilde\reg_k(\nu)-\reg(\nu) \right]$,
where $\tilde\reg_k(\cdot)$ is the approximation of $\reg_k(\cdot)$ around
the final estimate of $\coef_k$. For the case of $q=\moe$, the expression
obtained is (see Appendix)
\[
\zeta_\moe(\coef_k,\kappa,\beta)=
E_{\nu \sim \moe(\kappa,\beta)}\left[ \tilde\reg_k(\nu)-\reg(\nu) \right] = 
  \log(\coef_k+\beta) + \frac{1}{\coef_k+\beta}\left[\coef_k +
  \frac{\beta}{\kappa-1}\right] - \log \beta - \frac{1}{\kappa}.
\]
In the \moe case, for $\kappa$ and $\beta$ fixed, the minimum of
$\zeta_\moe$ occurs when
$\coef_k=\frac{\beta}{\kappa-1}=\mu(\beta,\kappa)$. We also have
$\zeta_\moe(0)=(\kappa-1)^{-1}-\kappa^{-1}$. 

The function $\zeta_q(\cdot)$ can be evaluated on each coefficient of
$\coefm$ to give an idea of its quality. For example, in the experiments
from Section~\ref{sec:results}, we obtained an average value of $0.16$,
which lies between $\zeta_\moe(0)=0.19$ and
$\min_\coef\zeta_\moe(a)=0.09$. Depending on the experiment, this represents
6\% to 7\% of the total sparse coding cost function value, showing the
efficiency of the proposed optimization.

{\bf Comments on parameter estimation:}\label{sec:universal:parameter-estimation} %
All the universal models presented so far, with the exception of the
conditional Jeffreys, depend on hyper-parameters which in principle should
be tuned for optimal performance (remember that they do not influence the
universality of the model). If tuning is needed, it is important to remember
that the proposed universal models are intended for reconstruction
coefficients of \emph{clean data}, and thus their hyper-parameters should be
computed from statistics of clean data, or either by compensating the
distortion in the statistics caused by noise (see for example
\cite{ramirez10dude}). Finally, note that when $\dictm$ is linearly
dependent and $\mathrm{rank}(\dictm) = \reals^\ndims$, the coefficients
matrix $\coefm$ resulting from an exact reconstruction of $\datam$ will have
many zeroes which are not properly explained by any continuous distribution
such as a Laplacian. We sidestep this issue by computing the
statistics only from the non-zero coefficients in $\coefm$. Dealing properly
with the case $P(\coef=0)>0$ is beyond the scope of this work.

\section{Experimental results}
%
\label{sec:results}

In the following experiments, the testing data $\datam$ are
$8{\times}8$ patches drawn from the Pascal VOC2006 \emph{testing}
subset,\footnote{\url{http://pascallin.ecs.soton.ac.uk/challenges/VOC/databases.html\#VOC2006}}
which are high quality $640{\times}480$ \textsc{rgb} images with 8 bits per
channel. For the experiments, we converted the 2600 images to grayscale by
averaging the channels, and scaled the dynamic range to lie in the $[0,1]$
interval. Similar results to those shown here are also obtained for other
patch sizes. 

\subsection{Dictionary learning}
\label{sec:results:learning}

For the experiments that follow, unless otherwise stated, we use a
``global'' overcomplete dictionary $\dictm$ with $\natoms=4\ndims=256$ atoms
trained on the full VOC2006 \emph{training} subset using the method
described in \cite{ramirez09aruba,ramirez10a}, which seeks to minimize the
following cost during training,\footnote{%
  While we could have used off-the-shelf dictionaries such as \textsc{dct}
  in order to test our universal sparse coding framework, it is important to
  use dictionaries that lead to the state-of-the-art results in order to
  show the additional potential improvement of our proposed regularizers.}
\begin{equation}
  \label{eq:incoherent-sparse-modeling}
\min_{\dictm,\coefm} \frac{1}{\nsamples}\sum_{j=1}^{\nsamples}\left\{
\norm{\datav_j - \dictm\,\coefv_j}_2^2 + \lambda\reg(\coefv_j)
\right\} + \mu\norm{\dictm^T\dictm}_F^2,
\end{equation}
where $\norm{\cdot}_F$ denotes Frobenius norm. The additional term,
$\mu\norm{\dictm^T\dictm}_F^2$, encourages \emph{incoherence} in the learned
dictionary, that is, it forces the atoms to be as orthogonal as
possible. Dictionaries with lower coherence are well known to have several
theoretical advantages such as improved ability to recover sparse signals
\cite{daubechies04,tropp04}, and faster and better convergence to the
solution of the sparse coding problems \refeq{eq:sparse-modeling} and
\refeq{eq:sparse-modeling-lagrangian} \cite{elad07b}. Furthermore, in
\cite{ramirez09aruba} it was shown that adding incoherence leads to
improvements in a variety of sparse modeling applications, including the
ones discussed below.

We used \moe as the regularizer in \refeq{eq:incoherent-sparse-modeling},
with $\lambda=0.1$ and $\mu=1$, both chosen empirically. See
~\cite{aharon06,mairal08e,ramirez09aruba} for details on the optimization of
\refeq{eq:sparse-modeling-lagrangian} and \refeq{eq:incoherent-sparse-modeling}.

\subsection{\moe\ as a prior for sparse coding coefficients}
\label{sec:results:fitting}

We begin by comparing the performance of the Laplacian and \moe priors for
fitting a single global distribution to the whole matrix $\coefm$. We
compute $\coefm$ using \refeq{eq:sparse-modeling} with $\epsilon\approx 0$
and then, following the discussion in
Section~\ref{sec:universal:parameter-estimation}, restrict our study to the
nonzero elements of $\coefm$.

The empirical distribution of $\coefm$ is plotted in
Figure~\ref{fig:fitting-and-denoising}(a), along with the best fitting
Laplacian, \moe, \joe, and a particularly good example of the conditional
Jeffreys (\textsc{cmoe}) distributions.\footnote{To compute the empirical
  distribution, we quantized the elements of $\coefm$ uniformly in steps of
  $2^{-8}$, which for the amount of data available, gives us enough detail
  and at the same time reliable statistics for all the quantized values.}
The \mle for the Laplacian fit is
$\hat\theta=\nsamples_1/\norm{\coefm}_1=27.2$ (here $\nsamples_1$ is the
number of nonzero elements in $\coefm$).  For \moe, using
\refeq{eq:moe-parameter-estimation}, we obtained $\kappa=2.8$ and
$\beta=0.07$. For \joe, $\tmin=2.4$ and $\tmax=371.4$. According to the
discussion in Section~\ref{sec:universal:cond-jeffreys}, we used the value
$\kappa=2.8$ obtained using the method of moments for \moe as a hint for
choosing $n_0=2$ ($\kappa_0=n_0+1=3\approx 2.8$), yielding $\beta_0=0.07$,
which coincides with the $\beta$ obtained using the method of moments. As
observed in Figure~\ref{fig:fitting-and-denoising}(a), in all cases the
proposed mixture models fit the data better, significantly better for both
Gamma-based mixtures, \moe\ and \cmoe, and slightly better for \joe. This is
further confirmed by the Kullback-Leibler divergence (\textsc{kld}) obtained
in each case. Note that \joe\ fails to significantly improve on the
Laplacian mode due to the excessively large estimated range
$[\tmin,\tmax]$. In this sense, it is clear that the \joe\ model is very
sensitive to its hyper-parameters, and a better and more robust estimation
would be needed for it to be useful in practice.

Given these results, hereafter we concentrate on the best case which is the
\moe\ prior (which, as detailed above, can be derived from the conditional
Jeffreys as well, thus representing both approaches).

From Figure~\ref{fig:many-laplacians}(e) we know that the optimal
$\hat\theta$ varies locally across different regions, thus, we expect the
mixture models to perform well also on a per-atom basis. This is confirmed
in Figure~\ref{fig:fitting-and-denoising}(b), where we show, for each row
$\coefv^k, k=1,\ldots,\natoms$, the difference in \textsc{kld} between the
globally fitted \moe\ distribution and the best per-atom fitted \moe, the
globally fitted Laplacian, and the per-atom fitted Laplacians
respectively. As can be observed, the \textsc{kld} obtained with the
\emph{global} \moe\ is significantly smaller than the global Laplacian in
all cases, and even the \emph{per-atom} Laplacians in most of the cases.
This shows that \moe, with only two parameters (which can be easily
estimated, as detailed in the text), is a much better model than $\natoms$
Laplacians (requiring $\natoms$ critical parameters) fitted specifically to
the coefficients associated to each atom. Whether these modeling
improvements have a practical impact is explored in the next experiments.

\subsection{Recovery of noisy sparse signals}
\label{sec:active-set-recovery}

Here we compare the active set recovery properties of the \moe prior, with
those of the $\cost{1}$-based one, on data for which the sparsity assumption
$|\actset_{j}| \leq L$ holds exactly for all $j$. To this end, we obtain
sparse approximations to each sample $\datav\col{j}$ using the
\cost{0}-based Orthogonal Matching Pursuit algorithm (\omp) on $\dictm$
\cite{mallat93}, and record the resulting active sets $\actset_{j}$ as
ground truth. The data is then contaminated with additive Gaussian noise of
variance $\sigma$ and the recovery is performed by solving
\refeq{eq:sparse-modeling} for $\coefm$ with $\epsilon=C\ndims\sigma^2$ and
either the \cost{1} or the \moe-based regularizer for $\reg(\cdot)$. We use
$C=1.32$, which is a standard value in denoising applications (see for
example \cite{mairal08b}).

\begin{figure*}
  \begin{center}%
    \includegraphics[width=\textwidth]{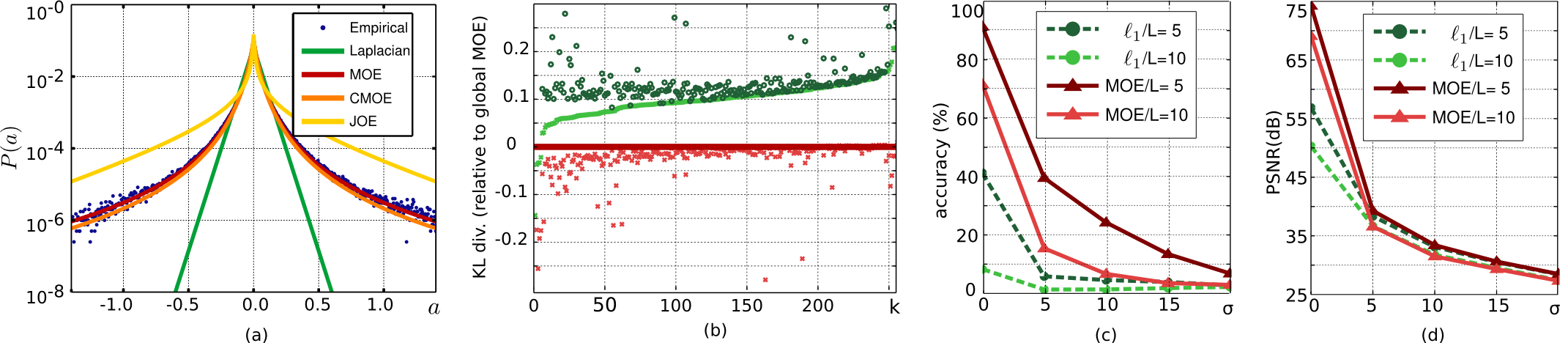}%
    \caption{\label{fig:fitting-and-denoising}\footnotesize%
      (a) Empirical distribution of the coefficients in $\coefm$ for image
      patches (blue dots), best fitting Laplacian (green), \moe (red), \cmoe
      (orange) and \joe (yellow) distributions. The Laplacian
      (\textsc{kld}$=\!0.17$ bits) is clearly not fitting the tails
      properly, and is not sufficiently peaked at zero either. The two
      models based on a Gamma prior, \moe (\textsc{kld}$=\!0.01$ bits) and
      \cmoe (\textsc{kld}$=\!0.01$ bits), provide an almost perfect fit. The
      fitted \joe (\textsc{kld}$=\!0.14$) is the most sharply peaked at $0$,
      but doest not fit the tails as tight as desired. As a reference, the
      entropy of the empirical distribution is $H=3.00$ bits.  (b)
      \textsc{kld} for the best fitting global Laplacian (dark green),
      per-atom Laplacian (light green), global \moe (dark red) and per-atom
      \moe (light red), relative to the \textsc{kld} between the globally
      fitted \moe distribution and the empirical distribution. The
      horizontal axis represents the indexes of each atom,
      $k=1,\ldots,\natoms$, ordered according to the difference in
      \textsc{kld} between the global \moe and the per-atom Laplacian model.
      Note how the global \moe outperforms both the global and per-atom
      Laplacian models in all but the first $4$ cases.
      (c) active set recovery accuracy of
       \cost{1} and \moe, as defined in
      Section~\ref{sec:active-set-recovery}, for $L=5$ and $L=10$, 
      as a function of $\sigma$. The improvement of \moe 
      over $\cost{1}$ is a factor of $5$ to $9$.
      (d) \psnr\ of the recovered sparse signals with respect to the true
      signals. In this case significant improvements can be observed at the
      high \textsc{snr} range, specially for highly sparse ($L=5$)
      signals. The performance of both methods is practically the same for
      $\sigma\geq10$. %
    }
\end{center}
\end{figure*}

For each sample $j$, we measure the error of each method in recovering the
active set as the Hamming distance between the true and estimated support of
the corresponding reconstruction coefficients. The accuracy of the method is
then given as the percentage of the samples for which this error falls below
a certain threshold $T$. Results are shown in
Figure~\ref{fig:fitting-and-denoising}(c) for $L=(5,10)$ and $T=(2,4)$
respectively, for various values of $\sigma$. Note the very significant
 improvement obtained with the proposed model.

Given the estimated active set $\actset_{j}$, the estimated clean patch
 is obtained by projecting $\datav\col{j}$ onto the
subspace defined by the atoms that are active according to $\actset_{j}$,
using least squares (which is the standard procedure for denoising once the
active set is determined). We then measure the \psnr of the estimated
patches with respect to the true ones. The results are shown in
Figure~\ref{fig:fitting-and-denoising}(d), again for various values of $\sigma$.
As can be observed, the \moe-based recovery is significantly better,
specially in the high \textsc{snr} range. Notoriously, the more accurate
active set recovery of \moe does not seem to improve the denoising
performance in this case. However, as we will see next, it does make a
difference when denoising real life signals, as well as for classification
tasks.

\subsection{Recovery of real signals with simulated noise}
\label{sec:unconstr-sparse-codi}

This experiment is an analogue to the previous one, when the data are the
original natural image patches (without forcing exact sparsity). Since for
this case the sparsity assumption is only approximate, and no ground truth
is available for the active sets, we compare the different methods in terms
of their denoising performance.

A critical strategy in image denoising is the use of overlapping patches,
where for each pixel in the image a patch is extracted with that pixel as
its center. The patches are denoised independently as $\ndims$-dimensional
signals and then recombined into the final denoised images by simple
averaging. Although this consistently improves the final result in all
cases, the improvement is very different depending on the method used to
denoise the individual patches. Therefore, we now compare the denoising
performance of each method at two levels: individual patches and final
image.

\begin{figure*}
\begin{center}
  \includegraphics[width=6.5in]{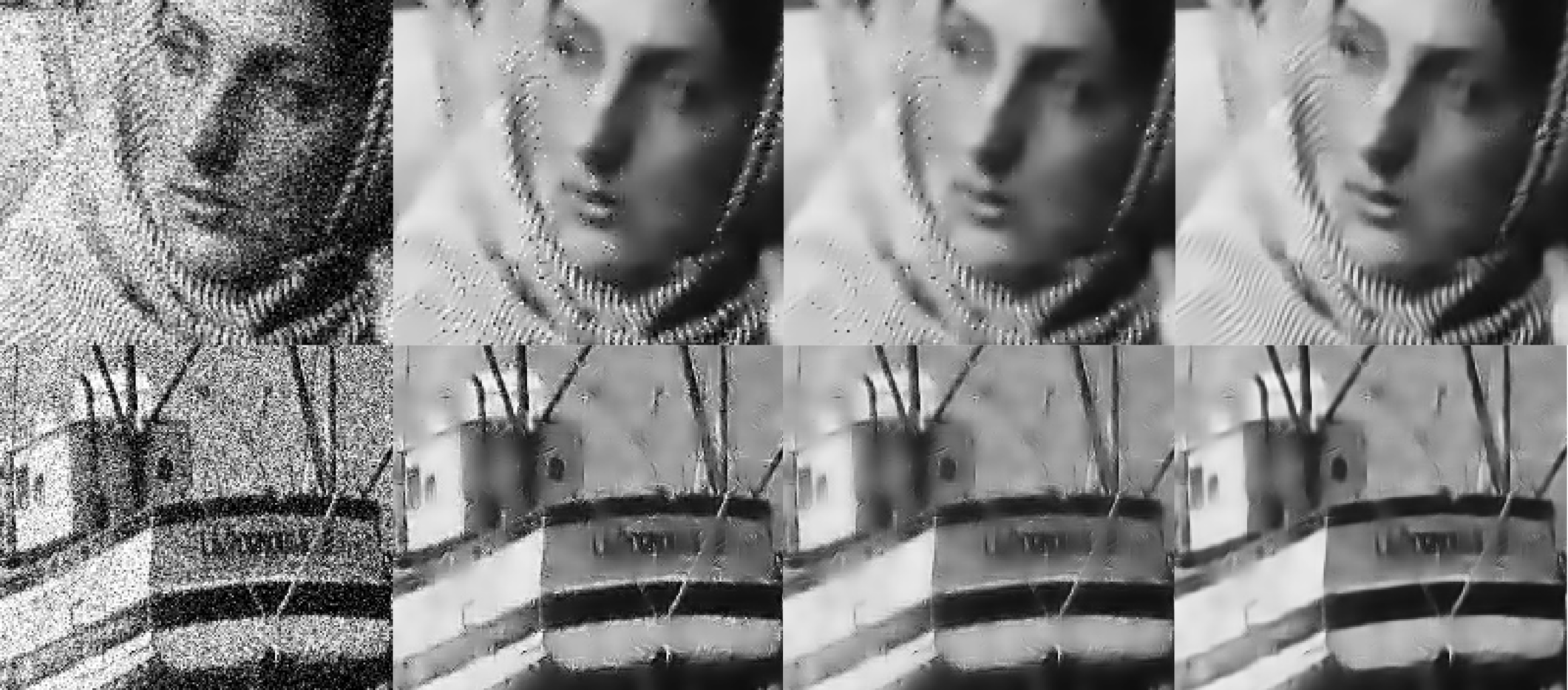}%
  \caption{\label{fig:denoising-sample}\footnotesize%
    Sample image denoising results. Top: Barbara, $\sigma=30$. Bottom:
    Boats, $\sigma=40$. From left to right: noisy, \cost{1}/\textsc{omp},
    \cost{1}/\cost{1}, \moe/\moe. The reconstruction obtained with the
    proposed model is more accurate, as evidenced by a better
    reconstruction of the texture in Barbara, and sharp edges in Boats,
    and does not produce the artifacts seen in both the \cost{1} and
    \cost{0} reconstructions, which appear as black/white speckles all over
    Barbara, and ringing on the edges in Boats.}
\end{center}
\end{figure*}

To denoise each image, the global dictionary described in
Section~\ref{sec:results:learning} is further adapted to the noisy image
patches using (\ref{eq:incoherent-sparse-modeling}) for a few iterations,
and used to encode the noisy patches via \refeq{eq:sparse-coding} with
$\epsilon=C\ndims\sigma^2$. We repeated the experiment for two learning
variants ($\cost{1}$ and \moe regularizers), and two coding variants
(\refeq{eq:sparse-coding} with the regularizer used for learning, and
\cost{0} via \textsc{omp}. The four variants were applied to the standard
images Barbara, Boats, Lena, Man and Peppers, and the results summarized in
Table~\ref{tab:denoising}. We show sample results in
Figure~\ref{fig:denoising-sample}. Although the quantitative improvements
seen in Table~\ref{tab:denoising} are small compared to \cost{1}, there is a
significant improvement at the visual level, as can be seen in
Figure~\ref{fig:denoising-sample}. In all cases the PSNR obtained coincides
or surpasses the ones reported in \cite{aharon06}.\footnote{Note that in
  \cite{aharon06}, the denoised image is finally blended with the noisy
  image using an empirical weight, providing an extra improvement to the
  final PSNR in some cases. The results in \ref{tab:denoising} are already
  better without this extra step.}

\newsavebox{\denoisingResultsTable}
\savebox{\denoisingResultsTable}{%
\begin{tabular}{|l|cc|cc|c|}\hline
%
%
	&\multicolumn{5}{|c|}{$\sigma=10$} \\ \hline
learning & \multicolumn{2}{|c}{\cost{1}} & \multicolumn{2}{|c|}{\textsc{moe}} & \cite{aharon06} \\
coding   & \cost{0} & \cost{1} &  \cost{0} & \textsc{moe} & \\ \hline
   barbara & 30.4/\best{34.4} & \best{31.2}/33.8 & 30.5/\best{34.4} & 30.9/\best{34.4} & \best{34.4} \\
      boat & 30.4/33.7 & \best{30.9}/33.4 & 30.5/33.7 & 30.8/\best{33.8} & 33.7 \\
      lena & 31.8/35.5 & \best{32.4}/35.1 & 32.1/35.6 & 32.3/\best{35.6} & 35.5 \\
   peppers & 31.6/34.8 & \best{32.1}/34.6 & 31.8/\best{34.9} & 32.0/\best{34.9} & 34.8 \\
       man & 29.6/33.0 & \best{30.6}/32.9 & 29.7/33.0 & 30.2/\best{33.1} & 32.8  \\
 AVERAGE   & 30.7/34.2 & \best{31.4}/33.9 & 30.8/34.2 & 31.1/\best{34.3} & 34.1  \\ \hline
%
%
	&\multicolumn{5}{|c|}{$\sigma=20$} \\ \hline
learning & \multicolumn{2}{|c}{\cost{1}} & \multicolumn{2}{|c|}{\textsc{moe}} & \cite{aharon06} \\
coding  & \cost{0} & \cost{1} &  \cost{0} & \textsc{moe} & \\ \hline
   barbara & 26.5/30.6 & 26.9/30.2 & 26.8/30.7 & \best{27.0}/\best{30.9} & 30.8 \\
      boat & 26.9/30.2 & 27.2/30.1 & 27.1/30.3 & \best{27.3}/\best{30.4} & 30.3 \\
      lena & 28.3/32.3 & 28.6/32.0 & 28.7/32.3 & \best{28.8}/\best{32.4} & \best{32.4} \\
   peppers & 28.3/31.9 & \best{28.7}/31.8 & 28.6/31.9 & \best{28.7}/\best{32.0} & 31.9 \\
       man & 25.8/28.8 & \best{26.3}/28.9 & 26.0/28.9 & 26.2/\best{29.0} & 28.8 \\
 AVERAGE   & 27.0/30.6 & 27.4/30.4 & 27.3/30.6 & \best{27.5}/\best{30.8} & 30.6 \\ \hline
\end{tabular}\hspace{0.25em}%
%
%
\begin{tabular}{|cc|cc|c|}\hline
\multicolumn{5}{|c|}{$\sigma=30$}  \\\hline
\multicolumn{2}{|c}{\cost{1}} & \multicolumn{2}{|c|}{\textsc{moe}} & \cite{aharon06} \\
\cost{0} & \cost{1} &  \cost{0} & \textsc{moe} & \\ \hline
 24.5/28.2 & 24.8/28.2 & 24.8/28.3 & \best{24.9}/\best{28.5} & 28.4 \\
 25.0/28.1 & 25.2/28.2 & 25.3/28.2 & \best{25.4}/\best{28.3} & 28.2 \\
 26.4/30.1 & 26.6/30.2 & 26.7/30.3 & \best{26.8}/\best{30.4} & 30.3 \\
 26.3/29.8 & 26.6/\best{29.9} & 26.6/\best{29.9} & \best{26.7}/\best{29.9} & - \\
 23.9/26.5 & \best{24.2}/\best{26.8} & 24.1/26.6 & \best{24.2}/26.7 & 26.5 \\
 25.1/28.3 & 25.4/\best{28.5} & 25.4/28.4 & \best{25.5}/\best{28.5} & 28.4 \\ \hline
%
%
\multicolumn{5}{|c|}{$\sigma=40$} \\ \hline
\multicolumn{2}{|c}{\cost{1}} & \multicolumn{2}{|c|}{\textsc{moe}} & \cite{aharon06} \\
\cost{0} & \cost{1} &  \cost{0} & \textsc{moe} & \\ \hline
 23.2/26.2 & 23.2/\best{26.5} & \best{23.4}/26.3 & \best{23.4}/26.4 & 26.4 \\
 23.7/26.4 & 23.8/\best{26.7} & \best{24.0}/26.6 & \best{24.0}/\best{26.7} & 26.7 \\
 25.1/28.5 & 25.2/\best{28.7} & \best{25.5}/28.6 & \best{25.5}/\best{28.7} & 28.7 \\
 24.9/27.9 & 25.1/\best{28.1} & 25.2/28.0 & \best{25.3}/\best{28.1} & 28.0 \\
 22.6/24.7 & 22.7/\best{25.0} & \best{22.8}/24.8 & \best{22.8}/24.9 & 24.8 \\
 23.8/26.5 & 23.9/\best{26.8} & \best{24.1}/26.6 & \best{24.1}/26.7 & 26.7 \\ \hline
\end{tabular}%
}

\savebox{\denoisingResultsTable}{%
\begin{tabular}{|l|cc|cc|c|}\hline
%
%
	&\multicolumn{5}{|c|}{$\sigma=10$} \\ \hline
learning & \multicolumn{2}{|c}{\cost{1}} & \multicolumn{2}{|c|}{\textsc{moe}} & \cite{aharon06} \\
coding   & \cost{0} & \cost{1} &  \cost{0} & \textsc{moe} & \\ \hline
   barbara & 30.4/\best{34.4} & \best{31.2}/33.8 & 30.5/\best{34.4} & 30.9/\best{34.4} & \best{34.4} \\
      boat & 30.4/33.7 & \best{30.9}/33.4 & 30.5/33.7 & 30.8/\best{33.8} & 33.7 \\
      lena & 31.8/35.5 & \best{32.4}/35.1 & 32.1/35.6 & 32.3/\best{35.6} & 35.5 \\
   peppers & 31.6/34.8 & \best{32.1}/34.6 & 31.8/\best{34.9} & 32.0/\best{34.9} & 34.8 \\
       man & 29.6/33.0 & \best{30.6}/32.9 & 29.7/33.0 & 30.2/\best{33.1} & 32.8  \\
 AVERAGE   & 30.7/34.2 & \best{31.4}/33.9 & 30.8/34.2 & 31.1/\best{34.3} & 34.1  \\ \hline
\end{tabular}
\begin{tabular}{|cc|cc|c|}\hline
%
%
\multicolumn{5}{|c|}{$\sigma=20$} \\ \hline
\multicolumn{2}{|c}{\cost{1}} & \multicolumn{2}{|c|}{\textsc{moe}} & \cite{aharon06} \\
 \cost{0} & \cost{1} &  \cost{0} & \textsc{moe} & \\ \hline
 26.5/30.6 & 26.9/30.2 & 26.8/30.7 & \best{27.0}/\best{30.9} & 30.8 \\
 26.9/30.2 & 27.2/30.1 & 27.1/30.3 & \best{27.3}/\best{30.4} & \best{30.3} \\
 28.3/32.3 & 28.6/32.0 & 28.7/32.3 & \best{28.8}/\best{32.4} & \best{32.4} \\
 28.3/31.9 & \best{28.7}/31.8 & 28.6/31.9 & \best{28.7}/\best{32.0} & 31.9 \\
 25.8/28.8 & \best{26.3}/28.9 & 26.0/28.9 & 26.2/\best{29.0} & 28.8 \\
 27.0/30.6 & 27.4/30.4 & 27.3/30.6 & \best{27.5}/\best{30.8} & 30.6 \\ \hline
\end{tabular}\hspace{0.25em}%
%
%
\begin{tabular}{|cc|cc|c|}\hline
\multicolumn{5}{|c|}{$\sigma=30$}  \\\hline
\multicolumn{2}{|c}{\cost{1}} & \multicolumn{2}{|c|}{\textsc{moe}} & \cite{aharon06} \\
\cost{0} & \cost{1} &  \cost{0} & \textsc{moe} & \\ \hline
 24.5/28.2 & 24.8/28.2 & 24.8/28.3 & \best{24.9}/\best{28.5} & 28.4 \\
 25.0/28.1 & 25.2/28.2 & 25.3/28.2 & \best{25.4}/\best{28.3} & 28.2 \\
 26.4/30.1 & 26.6/30.2 & 26.7/30.3 & \best{26.8}/\best{30.4} & 30.3 \\
 26.3/29.8 & 26.6/\best{29.9} & 26.6/\best{29.9} & \best{26.7}/\best{29.9} & - \\
 23.9/26.5 & \best{24.2}/\best{26.8} & 24.1/26.6 & \best{24.2}/26.7 & 26.5 \\
 25.1/28.3 & 25.4/\best{28.5} & 25.4/28.4 & \best{25.5}/\best{28.5} & 28.4 \\ \hline
\end{tabular}%
}

\begin{table*}\begin{center}
    \setlength{\tabcolsep}{2mm} \resizebox{\textwidth}{!}{%
      \usebox{\denoisingResultsTable}%
    } 
    \caption{\label{tab:denoising}\footnotesize%
      Denoising results: in each table, each column shows the denoising
      performance of a learning+coding combination. Results are shown in
      pairs, where the left number is the \textsc{psnr} between the clean
      and recovered individual patches, and the right number is the
      \textsc{psnr} between the clean and recovered images. Best results are
      in bold. The proposed \moe produces better final results over both the
      \cost{0} and \cost{1} ones in all cases, and at patch level for all
      $\sigma>10$. Note that the average values reported are the PSNR of the
      average MSE, and not the PSNR average.}
\end{center}
\end{table*}

\begin{figure*}
\begin{minipage}[t]{0.25\textwidth}%
\centering%
\setlength{\tabcolsep}{1mm}%
\resizebox{\textwidth}{2in}{\begin{tabular}[b]{|l|cccc|}\hline
image        & cubic &  \cost{0}  &   \cost{1}  &  \moe  \\ \hline
  barbara & 25.0 & 25.6 & 25.5 & 25.6 \\
     boat & 28.9 & 29.8 & 29.8 & 29.9 \\
     lena & 32.7 & 33.8 & 33.8 & 33.9 \\
  peppers & 32.0 & 33.4 & 33.4 & 33.4 \\
      man & 28.4 & 29.4 & 29.3 & 29.4 \\
    tools & 21.0 & 22.3 & 22.2 & 22.3 \\ \hline
AVER      & 22.8 & 24.0 & 24.0 & 24.1 \\\hline

\end{tabular}%
}%
\end{minipage}%
\hspace{1mm}%
\begin{minipage}[t]{0.74\textwidth}%
\includegraphics[width=\textwidth,height=2in]{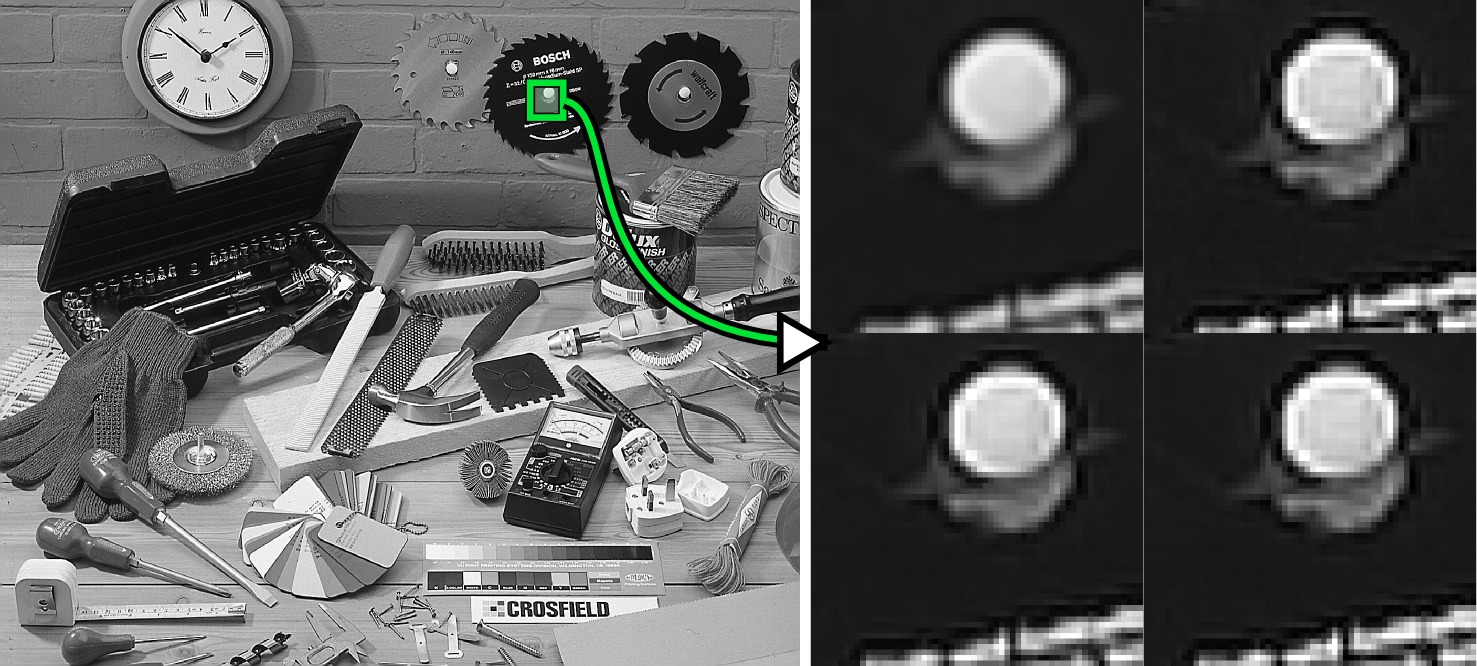}\\
\end{minipage}
\caption{\label{fig:zooming}Zooming results. Left to right:
  summary, Tools image, detail of zooming results for the framed region, top
  to bottom and left to right: cubic, \cost{0}, \cost{1}, \moe. As can be
  seen, the \moe result is as sharp as \cost{0} but produces less
  artifacts. This is reflected in the $0.1dB$ overall improvement obtained
  with \moe, as seen in the leftmost summary table. }
\end{figure*}

\subsection{Zooming}

As an example of signal recovery in the absence of noise, we took the
previous set of images, plus a particularly challenging one (Tools), and
subsampled them to half each side. We then simulated a zooming effect by
upsampling them and estimating each of the 75\% missing pixels (see e.g.,
\cite{guoshen10} and references therein).  We use a technique similar to the
one used in \cite{neelamani04}. The image is first interpolated and then
deconvoluted using a Wiener filter. The deconvoluted image has artifacts
that we treat as noise in the reconstruction. However, since there is no
real noise, we do not perform averaging of the patches, using only the
center pixel of $\hat\datav_j$ to fill in the missing pixel at $j$.  The
results are summarized in Figure~\ref{fig:zooming}, where we again observe
that using \moe instead of \cost{0} and \cost{1} improves the results.

\subsection{Classification with universal sparse models}

\newcommand{\nclasses}{c}
\newcommand{\class}{\mathcal{C}}
\newcommand{\energy}{\mathcal{R}}

In this section we apply our proposed universal models to a classification
problem where each sample $\datav\col{j}$ is to be assigned a class label
$y\col{j}=1,\ldots,\nclasses$, which serves as an index to the set of
possible classes, $\setdef{\class_1,\class_2,\ldots,\class_\nclasses}$.  We
follow the procedure of \cite{ramirez10a}, where the classifier assigns each
sample $\datav\col{j}$ by means of the maximum a posteriori criterion
\refeq{eq:map-gen} with the term $-\log P(\coefv)$ corresponding to the
assumed prior, and the dictionaries representing each class are learned from
training samples using \refeq{eq:incoherent-sparse-modeling} with the
corresponding regularizer $\reg(\coefv) = -\log P(\coefv)$. Each experiment
  is repeated for the baseline Laplacian model, implied in the \cost{1}
  regularizer, and the universal model \moe, and the results are then
  compared. In this case we expect that the more accurate prior model for
  the coefficients will result in an improved likelihood estimation, which
  in turn should improve the accuracy of the system.

\begin{figure*}
\includegraphics[width=0.125\textwidth]{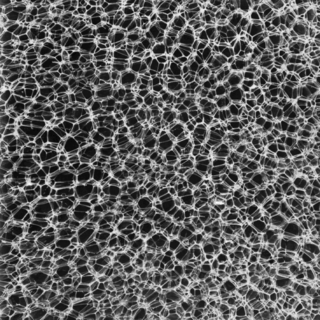}%
\includegraphics[width=0.125\textwidth]{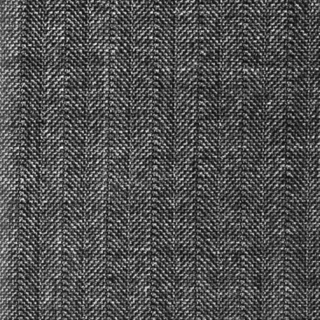}%
\includegraphics[width=0.125\textwidth]{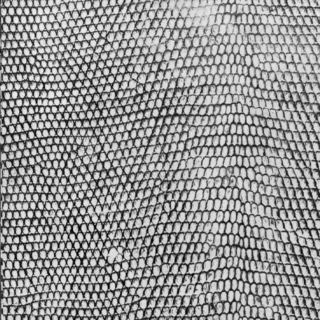}%
\includegraphics[width=0.125\textwidth]{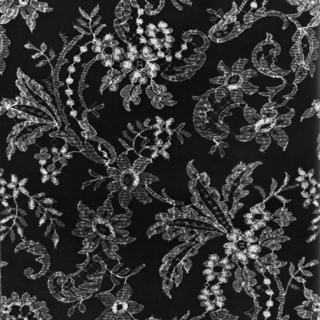}%
\includegraphics[width=0.125\textwidth]{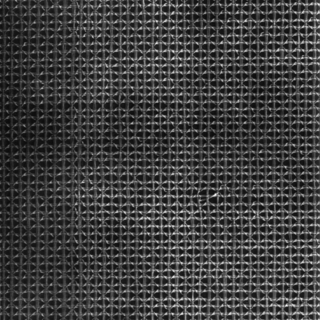}%
\includegraphics[width=0.125\textwidth]{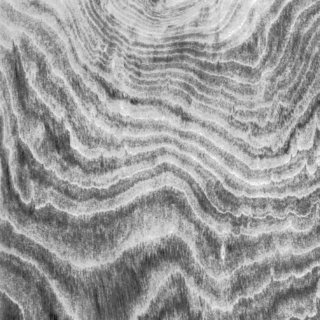}%
\includegraphics[width=0.125\textwidth]{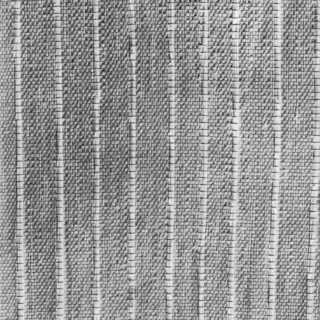}%
\includegraphics[width=0.125\textwidth]{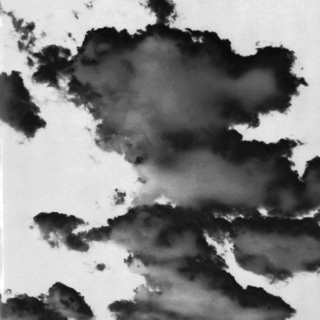}\\
\includegraphics[width=0.125\textwidth]{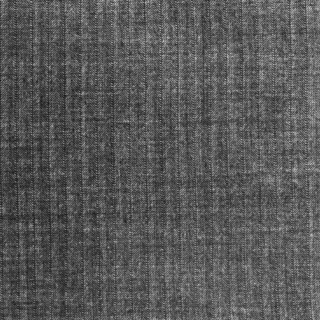}%
\includegraphics[width=0.125\textwidth]{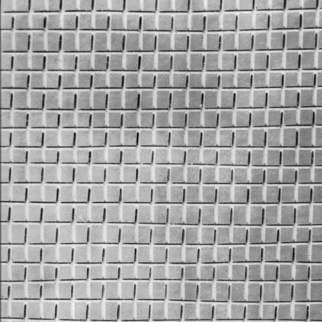}%
\includegraphics[width=0.125\textwidth]{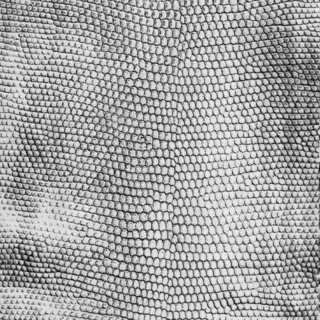}%
\includegraphics[width=0.125\textwidth]{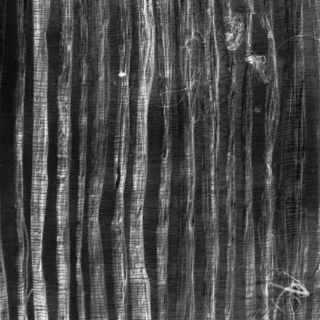}%
\includegraphics[width=0.125\textwidth]{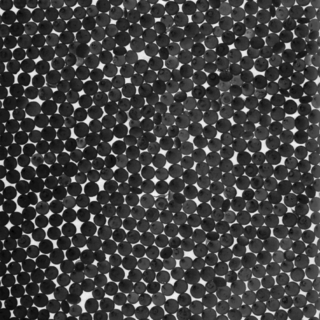}%
\includegraphics[width=0.125\textwidth]{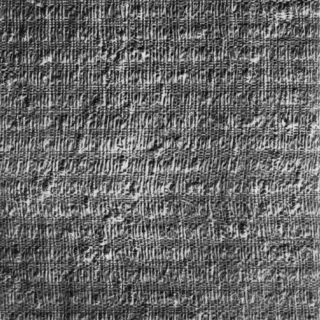}%
\includegraphics[width=0.125\textwidth]{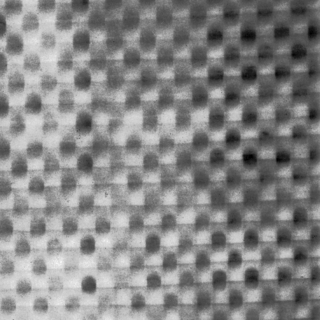}%
\includegraphics[width=0.125\textwidth]{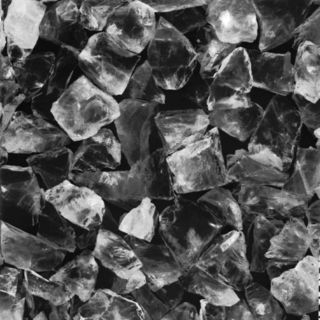}%
\caption{\label{fig:brodatz-subsample}Textures used in the texture
  classification example.}
\end{figure*}

  We begin with a classic texture classification problem, where patches have
  to be identified as belonging to one out of a number of possible
  textures. In this case we experimented with samples of $\nclasses=2$ and
  $\nclasses=3$ textures drawn at random from the Brodatz
  database,\footnote{\url{http://www.ux.uis.no/~tranden/brodatz.html}}, the
  ones actually used shown in Figure~\ref{fig:brodatz-subsample}. In each
  case the experiment was repeated $10$ times. In each repetition, a
  dictionary of $\natoms=300$ atoms was learned from all $16{\times}16$
  patches of the leftmost halves of each sample texture. We then classified
  the patches from the rightmost halves of the texture samples. For the
  $\nclasses=2$ we obtained an average error rate of $5.13\%$ using \cost{1}
  against $4.12\%$ when using \moe, which represents a reduction of $20\%$
  in classification error. For $\nclasses=3$ the average error rate obtained
  was $13.54\%$ using \cost{1} and $11.48\%$ using \moe, which is $15\%$
  lower. Thus, using the universal model instead of \cost{1} yields a
  significant improvement in this case (see for example \cite{mairal08e} for
  other results in classification of Brodatz textures).

  The second sample problem presented is the Graz'02 bike detection
  problem,\footnote{\url{http://lear.inrialpes.fr/people/marszalek/data/ig02/}}
  where each pixel of each testing image has to be classified as either
  background or as part of a bike. In the Graz'02 dataset, each of the
  pixels can belong to one of two classes: bike or background. On each of
  the training images (which by convention are the first 150 even-numbered
  images), we are given a mask that tells us whether each pixel belongs to a
  bike or to the background. We then train a dictionary for bike patches and
  another for background patches. Patches that contain pixels from both
  classes are assigned to the class corresponding to the majority of their
  pixels.

\begin{figure*}
  \begin{center}
    \includegraphics[width=6.5in]{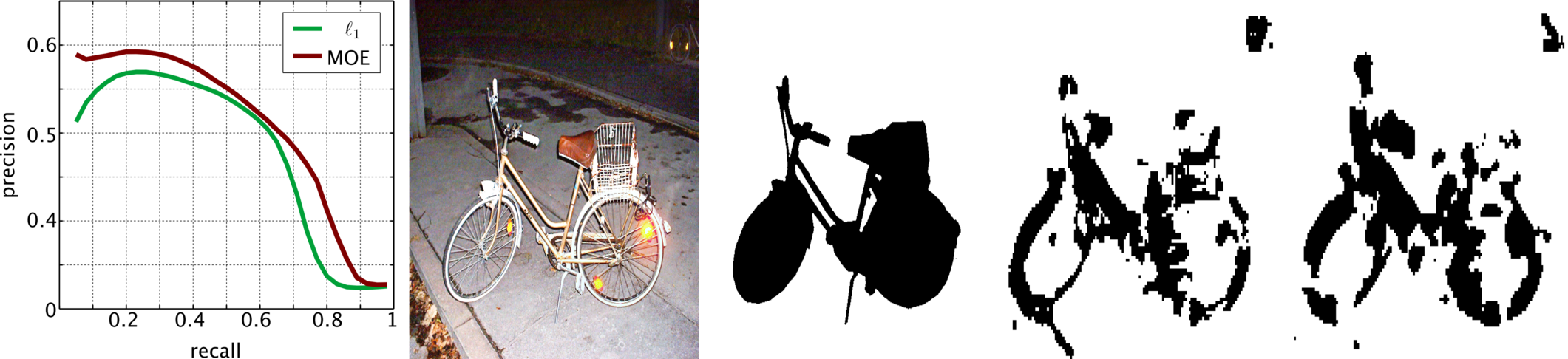}%
    \caption{\label{fig:classification}\footnotesize%
      Classification results. Left to right: precision vs. recall curve, a
      sample image from the Graz'02 dataset, its ground truth, and the
      corresponding estimated maps obtained with \cost{1} and \moe for a
      fixed threshold.  The precision vs. recall curve shows that the
      mixture model gives a better precision in all cases. In the example,
      the classification obtained with \moe\ yields less false positives and
      more true positives than the one obtained with \cost{1}.}
  \end{center}
\end{figure*}

In Figure~\ref{fig:classification} we show the \emph{precision vs. recall
  curves} obtained with the detection framework when either the \cost{1} or
the \moe regularizers were used in the system.  As can be seen, the \moe-based model outperforms the \cost{1}
in this classification task as well, giving a better precision for all
recall values.

In the above experiments, the parameters for the $\cost{1}$ prior
($\lambda$), the \moe model ($\lambda_{\moe}$) and the incoherence term
($\mu$) were all adjusted by cross validation. The only exception is the
\moe parameter $\beta$, which was chosen based on the fitting experiment as
$\beta=0.07$. 

\section{Concluding remarks}
\label{sec:conclusions}

A framework for designing sparse modeling priors was introduced in this
work, using tools from universal coding, which formalizes sparse coding and
modeling from a MDL perspective. The priors obtained lead to models with
both theoretical and practical advantages over the traditional \cost{0} and
\cost{1}-based ones. In all derived cases, the designed non-convex problems
are suitable to be efficiently (approximately) solved via a few iterations
of (weighted) \cost{1} subproblems. We also showed that these priors are
able to fit the empirical distribution of sparse codes of image patches
significantly better than the traditional \textsc{iid} Laplacian model, and
even the non-identically distributed independent Laplacian model where a
different Laplacian parameter is adjusted to the coefficients associated to
each atom, thus showing the flexibility and accuracy of these proposed
models. The additional flexibility, furthermore, comes at a small cost of
only $2$ parameters that can be easily and efficiently tuned (either
$(\kappa,\beta)$ in the \moe model, or $(\tmin,\tmax)$ in the \joe model),
instead of $\natoms$ (dictionary size), as in weighted \cost{1} models. The
additional accuracy of the proposed models was shown to have significant
practical impact in active set recovery of sparse signals, image denoising,
and classification applications.  Compared to the Bayesian approach, we
avoid the potential burden of solving several sampled sparse problems, or
being forced to use a conjugate prior for computational reasons (although in
our case, {\it a fortiori}, the conjugate prior does provide us with a good
model). Overall, as demonstrated in this paper, the introduction of
information theory tools can lead to formally addressing critical aspects of
sparse modeling.

Future work in this direction includes the design of priors that take into
account the nonzero mass at $\coef=0$ that appears in overcomplete models,
and online learning of the model parameters from noisy data, following for
example the technique in \cite{ramirez10dude}.

\section*{Acknowledgments}

Work partially supported by \textsc{nga}, \textsc{onr}, \textsc{aro},
\textsc{nsf}, \textsc{nsseff}, and \textsc{fundaciba-antel}. We wish to
thank Julien Mairal for providing us with his fast sparse modeling toolbox,
SPAMS.\footnote{\url{http://www.di.ens.fr/willow/SPAMS/}} We also thank
Federico Lecumberry for his participation on the incoherent dictionary
learning method, and helpful comments.

\appendix

\section*{Derivation of the MOE model}
\label{sec:moe-details}

In this case we have $ P(\coef|\theta)={\theta}e^{-\theta \coef}$ and
$w(\theta|\kappa,\beta) =
\frac{1}{\Gamma(\kappa)}\theta^{\kappa-1}\beta^{\kappa}e^{-\beta\theta}, $
which, when plugged into \refeq{eq:scalar-mixture}, gives
\begin{eqnarray*}
  Q(\coef|\beta,\kappa) & = & 
  \int_{\theta=0}^{\infty}{
    \theta e^{-\theta \coef} 
    \frac{1}{\Gamma(\kappa)}\theta^{\kappa-1}\beta^{\kappa}e^{-\beta\theta}
    d\theta
   } 
   \;=\; 
  \frac{\beta^{\kappa}}{\Gamma(\kappa)}\int_{\theta=0}^{\infty}{
    e^{-\theta(\coef + \beta)} 
    \theta^{\kappa}d\theta.
   }
\end{eqnarray*}
After the change of variables $u:=(\coef+\beta)\theta$ ($u(0)=0$,
$u(\infty)=\infty$), the integral can be written as
\begin{eqnarray*}
  Q(\coef|\beta,\kappa) & = & 
  \frac{\beta^{\kappa}}{\Gamma(\kappa)}\int_{\theta=0}^{\infty}{
    e^{-u} \left(\frac{u}{\coef+\beta}\right)^k\frac{du}{\coef+\beta}
   } 
  \; = \; \frac{\beta^{\kappa}}{\Gamma(\kappa)}(\coef+\beta)^{-(\kappa+1)}
  \int_{\theta=0}^{\infty}{ e^{-u} u^k du } \\
  & = & \frac{\beta^{\kappa}}{\Gamma(\kappa)}(\coef+\beta)^{-(\kappa+1)}\Gamma(\kappa+1)
  \; = \; \frac{\beta^{\kappa}}{\Gamma(\kappa)}(\coef+\beta)^{-(\kappa+1)}\kappa\Gamma(\kappa),
\end{eqnarray*}
obtaining $Q(\coef|\beta,\kappa)  =  \kappa\beta^{\kappa}(\coef+\beta)^{-(\kappa+1)},$
since the integral on the second line is precisely the definition of
$\Gamma(\kappa+1)$. The symmetrization is obtained by substituting $\coef$
by $\abs{\coef}$ and dividing the normalization constant by two,
$Q(|\coef||\beta,\kappa)  = 0.5\kappa\beta^{\kappa}(|\coef|+\beta)^{-(\kappa+1)}.$

The mean of the \textsc{moe} distribution (which is defined only for
$\kappa>1$) can be easily computed using integration by parts,
\begin{eqnarray*}
\mu(\beta,\kappa) & = & \kappa\beta^\kappa\int_{0}^{\infty}{ \frac{u}{(u+\beta)^{(\kappa+1)}}du }
                  \; = \; \kappa\beta\left[-\left.\frac{u}{\kappa(u+\beta)^\kappa}\right|_{0}^{\infty} +
                        \frac{1}{\kappa}\int_{0}^{\infty}{ \frac{du}{(u+\beta)^k}  } \right] 
                   = \frac{\beta}{\kappa-1}.
\end{eqnarray*}
In the
same way, it is easy to see that the non-central moments of order $i$ are
$\mu_i = \frac{\beta}{{\kappa-1 \choose i}}.$

The \textsc{mle} estimates of $\kappa$ and $\beta$ can be obtained using any
nonlinear optimization technique such as Newton method, using for example
the estimates obtained with the method of moments as a starting point. In
practice, however, we have not observed any significant improvement in using
the \textsc{mle} estimates over the moments-based ones.


\subsection*{Expected approximation error in cost function}

As mentioned in the optimization section, the LLA approximates the \moe
regularizer as a weighted \cost{1}. Here we develop an expression for the
expected error between the true function and the approximate convex one,
where the expectation is taken (naturally) with respect to the \moe
distribution. Given the value of the current iterate
$\coef\iter{t}=\coef_0$, (assumed positive, since the function and its
approximation are symmetric), the approximated regularizer is
$\reg\iter{t}(\coef)=\log(\coef_0+\beta) +
\frac{1}{|\coef_0|+\beta}(\coef-\coef_0)$. We have
\begin{eqnarray*}
E_{\coef \sim \moe(\kappa,\beta)}\left[ \reg\iter{t}(\coef)-\reg(\coef) \right] & = &
\int_{0}^{\infty}{\frac{\kappa\beta^\kappa}{(\coef+\kappa)^{\kappa+1}} \left[
\log(|\coef_0+\beta) +
\frac{1}{\coef_0+\beta}(\coef-\coef_0) - \log(\coef + \beta)\right] d\coef} \\
\!\!\!\! &=& \log(\coef_0+\beta) + 
    \frac{\coef_0}{\coef_0+\beta} + 
    \frac{\kappa\beta^\kappa}{\coef_0 + \beta}\int_{0}^{\infty}{\frac{\coef}{(\coef+\beta)^{\kappa+1}}d\coef}-
    \kappa\beta^\kappa \int_0^\infty{\frac{\log(\coef+\beta)}{(\coef+\beta)^{\kappa+1}}d\coef} \\
\!\!\!\! &= &\log(\coef_0+\beta) + 
    \frac{\coef_0}{\coef_0+\beta} + 
    \frac{\beta}{(\coef_0+\beta)(\kappa-1)} - \log \beta - \frac{1}{\kappa}.
\end{eqnarray*}

\section*{Derivation of the constrained Jeffreys (JOE) model}
\label{sec:joe-details}

In the case of the exponential distribution, the Fisher Information Matrix
in \refeq{eq:fisher-information} evaluates to
\begin{eqnarray*}
I(\theta) &=& 
\left.\left\{ E_{P(\cdot|\tilde\theta)}
\left[\frac{\partial^2}{\partial\tilde\theta^2}(-\log\theta+\theta\log \coef)
  \right]\right\}\right|_{\tilde\theta=\theta} 
\;=\;
\left.\left\{ E_{P(\cdot|\tilde\theta)}
\left[\frac{1}{\tilde\theta^2}
\right]\right\}\right|_{\tilde\theta=\theta} = \frac{1}{\theta^2}.
\end{eqnarray*}

By plugging this result into \refeq{eq:jeffreys-prior} with 
$\Theta=[\tmin,\tmax]$, $0 < \tmin < \tmax < \infty$ we obtain
$w(\theta)=\frac{1}{\ln(\tmax/\tmin)}\frac{1}{\theta}.$ We now derive the
(one-sided) \textsc{joe} probability density function by plugging this
$w(\theta)$ in \refeq{eq:scalar-mixture},
\begin{eqnarray*}
Q(\coef) & = & \int_{\tmin}^{\tmax}{\theta e^{-\theta \coef} \frac{1}{\ln(\tmax/\tmin)}\frac{d\theta}{\theta}} 
              \; = \; \frac{1}{\ln(\tmax/\tmin)} \int_{\tmin}^{\tmax}{ e^{-\theta \coef} d\theta } 
              \; = \; \frac{1}{\ln(\tmax/\tmin)} \frac{1}{\coef} \left( e^{-\tmin\coef} -e^{-\tmax\coef} \right).
\end{eqnarray*}
Although $Q(\coef)$ cannot be evaluated at $\coef=0$, the limit for
$\coef \rightarrow 0$ exists and is finite, so we can just define $Q(0)$ as
this limit, which is
\begin{eqnarray*}
\lim_{\coef \rightarrow 0} Q(\coef) 
& = & 
\lim_{\coef \rightarrow 0} \frac{1}{\ln(\tmax/\tmin)\coef}\left[ 1-\tmin\coef+o(\coef^2) - (1-\tmax\coef+o(\coef^2)) \right] 
= \frac{\tmax-\tmin}{\ln(\tmax/\tmin)}.
\end{eqnarray*}

Again, if desired, parameter estimation can be done for example using
maximum likelihood (via nonlinear optimization), or using the method of
moments. However, in this case, the method of moments does not provide a
closed form solution for $(\tmin,\tmax)$. The non-central moments of order
$i$ are
\begin{equation}
\mu_i  =  \int_{0}^{\infty+}{\!\frac{\coef^i}{\ln(\tmax/\tmin)}\frac{1}{\coef}\left[e^{-\tmin\coef}-e^{-\tmin\coef}\right]d\coef} 
\!=\! \frac{1}{\ln(\tmax/\tmin)}\left\{
\int_{0}^{+\infty}{\!\coef^{i-1}e^{-\tmin\coef}d\coef} -
\int_{0}^{+\infty}{\!\coef^{i-1}e^{-\tmax\coef}d\coef}\right\}.
\label{eq:joe-moments}
\end{equation}
For $i=1$, both integrals in \refeq{eq:joe-moments} are trivially evaluated,
yielding $\mu_1 = \frac{1}{\ln(\tmax/\tmin)}(\tmin^{-1}-\tmax^{-1})$. For
$i>1$, these integrals can be solved using integration by parts:
\begin{eqnarray*}
\mu_i^+ &= &\int_{0}^{+\infty}{\coef^{i-1}e^{-\tmin\coef}d\coef}  = 
\left.\coef^{i-1}\frac{1}{(-\tmin)}e^{-\tmin\coef}\right|_0^{+\infty} -
\frac{1}{(-\tmin)}(i-1)\int_{0}^{+\infty}{\coef^{i-2}e^{-\tmin\coef}d\coef}  \\
\mu_i^- & = & \int_{0}^{+\infty}{\coef^{i-1}e^{-\tmax\coef}d\coef}  = 
\left.\coef^{i-1}\frac{1}{(-\tmax)}e^{-\tmax\coef}\right|_0^{+\infty} -
\frac{1}{(-\tmax)}(i-1)\int_{0}^{+\infty}{\coef^{i-2}e^{-\tmax\coef}d\coef},
\end{eqnarray*}
where the first term in the right hand side of both equations evaluates to
$0$ for $i>1$. Therefore, for $i>1$ we obtain the recursions
%
$\mu_i^+ = \frac{i-1}{\tmin}\mu_{i-1}^+,\;
\mu_i^- = \frac{i-1}{\tmax}\mu_{i-1}^-,$
which, combined with the result for $i=1$, give the final expression for
all the moments of order $i>0$
\[ 
\mu_i  =
\frac{(i-1)!}{\ln(\tmax/\tmin)}\left(\frac{1}{\tmin^i}-\frac{1}{\tmax^i}
\right),\;i=1,2,\ldots.
\] 
In particular, for $i=1$ and $i=2$ we have
$\tmin  =  \left( \ln(\tmax/\tmin)\mu_1 + \tmax^{-1} \right)^{-1},\;
\tmax  =  \left( \ln(\tmax/\tmin)\mu_2 + \tmin^{-2} \right)^{-1},$
which, when combined, give us
\begin{equation}
\tmin  =  \frac{2\mu_1}{\mu_2 + \ln(\tmax/\tmin)\mu_1^2},
\quad
\tmax  =  \frac{2\mu_1}{\mu_2 - \ln(\tmax/\tmin)\mu_1^2}.\label{eq:joe-method-of-moments}
\end{equation}

One possibility is to solve the nonlinear equation $\tmax/\tmin =
\frac{\mu_2 + \ln(\tmax/\tmin)\mu_1^2}{\mu_2 - \ln(\tmax/\tmin)\mu_1^2}$ for
$u=\tmin/\tmax$ by finding the roots of the nonlinear equation $u =
\frac{\mu_2 + \ln u\mu_1^2}{\mu_2 - \ln u\mu_1^2}$ and choosing one of them
based on some side information. Another possibility is to simply fix the
ratio $\tmax/\tmin$ beforehand and solve for $\tmin$ and $\tmax$ using
\refeq{eq:joe-method-of-moments}.

\section*{Derivation of the conditional Jeffreys (CMOE) model}
\label{sec:cmoe-details}

The conditional Jeffreys method defines a proper prior $w(\theta)$ by
assuming that $n_0$ samples from the data to be modeled $\coefv$ were
already observed. Plugging the Fisher information for the exponential distribution,
 $I(\theta)=\theta^{-2}$, into (\ref{eq:cmoe-gen}) we obtain
\begin{eqnarray*}
w(\theta) & = &
\frac{P(\coef^{n_0}|\theta)\theta^{-1}}{\int_{\Theta}{P(\coef^{n_0}|\xi)\xi^{-1}d\xi}} 
 \; = \;
\frac{(\prod_{j=1}^{n_0}\theta e^{-\theta \coef_j}) \theta^{-1}}
{\int_{0}^{+\infty}{(\prod_{j=1}^{n_0}\xi e^{-\xi \coef_j})\xi^{-1}d\xi}} 
=
\frac{\theta^{n_0-1} e^{-\theta \sum_{j=1}^{n_0}\coef_j}}
{\int_{0}^{+\infty}{\xi^{n_0-1} e^{-\xi \sum_{j=1}^{n_0}\coef_j}d\xi}}.
\end{eqnarray*}

Denoting $S_0 = \sum_{j=1}^{n_0}\coef_j$ and performing the change of
variables $u := S_0 \xi$ we obtain
\[
w(\theta)  = 
\frac{\theta^{n_0-1} e^{-S_0\theta}}
{S_0^{-n_0}\int_{0}^{+\infty}{u^{n_0-1} e^{-u}du}} \; = \;
\frac{S_0^{n_0}\theta^{n_0-1} e^{-S_0\theta}}
{\Gamma(n_0)},
\]
where the last equation derives from the definition of the Gamma function,
$\Gamma(n_0)$.  We see that the resulting prior $w(\theta)$ is a Gamma
distribution Gamma$(\kappa_0,\beta_0)$ with $\kappa_0=n_0$ and
$\beta_0=S_0=\sum_{j=1}^{n_0}\coef_j$.


\bibliographystyle{plain}
\end{document}